\documentclass[12pt]{article}
\usepackage{amssymb}
\usepackage{epsfig}

\newcommand\nn{\nonumber}
\newcommand\lt{\left}
\newcommand\rt{\right}

\begin{document}

\title{The Spectrum of Density Perturbations\\
produced during Inflation to Leading Order in\\
a General Slow-Roll Approximation}
\author{Ewan D. Stewart \footnote{ewan@kaist.ac.kr}\\
{\em Department of Physics, KAIST, Taejeon 305-701, South Korea}}
\maketitle

\begin{abstract}
The standard calculation of the spectrum of density perturbations
produced during inflation assumes
(i) $|n-1| \ll 1$ and (ii) $|dn/d\ln k| \ll |n-1|$.
Slow-roll predicts and observations require (i),
but neither slow-roll nor observations require (ii).
In this paper I derive formulae for the spectrum $\mathcal{P}$,
spectral index $n$, and running of the spectral index $dn/d\ln k$,
assuming only (i) and not (ii).
I give a large class of observationally viable examples in which these
general slow-roll formulae are accurate but the standard slow-roll
formulae for the spectral index and running are incorrect.
\end{abstract}

\thispagestyle{empty}
\setcounter{page}{0}
\newpage
\setcounter{page}{1}

\section{Introduction}

Inflation \cite{Gliner,Guth} explains why the universe is big, full of matter,
and approximately spatially homogeneous, isotropic and flat on the largest
observable scales.
It also produces curvature perturbations \cite{Starobinsky} which eventually
grow to produce all the structure in the observable universe.
These curvature perturbations also provide a unique observational
opportunity to determine the more detailed properties of inflation.
In this paper, I will show that the standard formulae relating the spectrum
of curvature perturbations to the properties of inflation, such as
\begin{equation}
n_{\mathcal{R}_\mathrm{c}}(k)
= \lt. 1 - 3 \lt(\frac{V'}{V}\rt)^2 + 2 \frac{V''}{V} \rt|_{aH=k}
\end{equation}
can be incorrect even in the context of slow-roll inflation and current
observational bounds,
and provide new formulae which are robust.
These formulae will clearly be important if one wants to use observations
to probe the properties of inflation in a model independent way.
Some of these formulae have already been presented and discussed in a
companion paper \cite{Scott}.
Here I give a complete set of formulae with derivations and more examples.

The outline of the paper is as follows.
Section~1: Introduction.
Section~2: Motivation.
Section~3: Formalism.
Section~4: Series Formulae for the Spectrum in terms of
the slow-roll parameters and in terms of the inflaton potential.
Section~5: Integral Formulae for the Spectrum in terms of
the slow-roll parameters and in terms of the inflaton potential.
Section~6: Optimized Standard Slow-Roll.
Section~7: Examples.
The Appendices contain a summary of notation and various secondary formulae
used in this paper, give the relationship between the series formulae and
the integral formulae, and give explicit formulae for the coefficients of the
series formulae.

\section{Motivation}

In slow-roll \cite{slowroll} inflation,
with a single component inflaton,
the energy density is dominated by the potential energy
\begin{equation}
\rho = \frac{1}{2} \dot{\phi}^2 + V \simeq V
\end{equation}
and the equation of motion
\begin{equation}
\ddot\phi + 3H\dot\phi + V' = 0
\end{equation}
reduces to the slow-roll equation of motion
\begin{equation}
3H\dot\phi + V' \simeq 0
\end{equation}
Using $\rho = 3H^2$, the first of these conditions can be expressed as
\begin{equation}
\epsilon \equiv \frac{1}{2} \lt(\frac{\dot\phi}{H}\rt)^2
= \mathcal{O}(\xi)
\end{equation}
and the second as
\begin{equation}
\delta_1 \equiv \frac{\ddot\phi}{H\dot\phi} = \mathcal{O}(\xi)
\end{equation}
for some small parameter $\xi$.
In this approximation, the spectrum of curvature perturbations
produced during inflation is
\begin{equation}\label{Psr}
\mathcal{P}_{\mathcal{R}_\mathrm{c}}(k)
= \lt(\frac{H}{2\pi}\rt)^2 \lt(\frac{H}{\dot\phi}\rt)^2
\lt[ 1 + \mathcal{O}(\xi) \rt]
\end{equation}
where the right hand side should be evaluated around the time the mode $k$
left the horizon during inflation.
Now
\begin{equation}\label{Hdot}
\frac{d\ln H}{d\ln a} = - \epsilon = \mathcal{O}(\xi)
\end{equation}
and
\begin{equation}\label{phidot}
\frac{d\ln\dot\phi}{d\ln a} = \delta_1 = \mathcal{O}(\xi)
\end{equation}
and so we see that the spectrum is approximately scale invariant
\begin{equation}
n_{\mathcal{R}_\mathrm{c}}
\equiv 1 + \frac{d \ln \mathcal{P}_{\mathcal{R}_\mathrm{c}}}{d\ln k}
= 1 + \mathcal{O}(\xi)
\end{equation}
as is required by observations, and also by the fact that we have not
specified an exact time to evaluate the right hand side of Eq.~(\ref{Psr}).
Thus slow-roll is well justified by observations.

To calculate the deviation of the spectral index from one,
the standard {\em extra\/} assumption is to assume that the
$\mathcal{O}(\xi)$ terms in Eq.~(\ref{Psr}) are also approximately scale
invariant and so do not contribute to the spectral index at order $\xi$.
With this extra assumption one gets
\begin{equation}\label{nsr}
n_{\mathcal{R}_\mathrm{c}}(k) - 1
= - 4 \epsilon - 2 \delta_1 + \mathcal{O}\lt(\xi^2\rt)
\end{equation}
where again the right hand side should be evaluated around the time the
mode $k$ left the horizon during inflation.
The $\mathcal{O}(\xi)$ terms in Eq.~(\ref{Psr}) involve $\epsilon$ and
$\delta_1$ and so Eq.~(\ref{nsr}) assumes $\epsilon$ and $\delta_1$
are approximately scale invariant, as can also be seen from the fact that
we have not specified an exact time to evaluate the right hand side of
Eq.~(\ref{nsr}).
This in turn implies
\begin{equation}\label{npsr}
\frac{d n_{\mathcal{R}_\mathrm{c}}}{d\ln k} = \mathcal{O}\lt(\xi^2\rt)
\end{equation}
which is {\em not\/} required by observations.
Thus the standard extra assumption used to derive Eq.~(\ref{nsr}),
\mbox{i.e.} to determine the deviation of the spectral index from one,
is not justified by observations, nor is it a consequence of slow-roll.

If this extra assumption is not correct then the standard formula for the
power spectrum, Eq.~(\ref{Psr}), will be correct to leading order but the
standard formulae for the spectral index, Eq.~(\ref{nsr}), will not be
correct even to leading order, while the standard estimate for the
running of the spectral index, Eq.~(\ref{npsr}), will not even be the
correct order.
This is because, for the spectral index to be close to 1 as is indicated
by observations, the leading term in the power spectrum should be
approximately scale invariant, but the subleading terms need not be
and so can contribute at leading order to the spectral index and
its derivatives.

In this paper I will derive formulae for the spectrum, spectral index and
running of the spectral index without making any extra assumptions beyond
slow-roll, which is well justified by observations.
In this general slow-roll approximation
$d n_{\mathcal{R}_\mathrm{c}}/d\ln k$ can be of the same order as
$n_{\mathcal{R}_\mathrm{c}} - 1$, rather than necessarily being of order
$(n_{\mathcal{R}_\mathrm{c}} - 1)^2$ as is the case with the standard
assumptions.

\section{Formalism}

I will mostly follow the formalism of Ref.~\cite{Jinook} and only very
briefly review it here.
Defining
\begin{equation}\label{z}
z \equiv \frac{a\dot \phi}{H}
\end{equation}
and
\begin{equation}
\varphi \equiv a\left(\delta\phi-\frac{\dot\phi}{H}\mathcal{R}\right)
= -z\mathcal{R}_\mathrm{c}
\end{equation}
the equation of motion for the Fourier modes of the scalar perturbations is
\cite{Mukhanov}
\begin{equation}
\frac{d^2\varphi_k}{d\eta^2}
+ \left(k^2-\frac{1}{z}\frac{d^2z}{d\eta^2}\right)\varphi_k = 0
\end{equation}
with the asymptotic conditions
\begin{equation}\label{bc}
\varphi_k \longrightarrow \left\{
\begin{array}{l l l}
\frac{1}{\sqrt{2k}}e^{-ik\eta} & \mbox{as} & -k\eta \rightarrow \infty \\
A_k z & \mbox{as} & -k\eta \rightarrow 0
\end{array} \right.
\end{equation}
The power spectrum is given by
\begin{equation}
\mathcal{P}_{\mathcal{R}_\mathrm{c}}
= \left(\frac{k^3}{2\pi^2}\right) \lim_{-k\eta \rightarrow 0} \left|\frac{\varphi_k}{z}\right|^2
= \frac{k^3}{2\pi^2}|A_k|^2
\end{equation}
Now define $y \equiv \sqrt{2k}\, \varphi_k$ and $x \equiv -k\eta$,
and choose the ansatz that $z$ takes the form
\begin{equation}\label{f}
z = \frac{1}{x} \, f(\ln x)
\end{equation}
Then the equation of motion is
\begin{equation}\label{yeq}
\frac{d^2y}{dx^2} + \left(1-\frac{2}{x^2}\right) y = \frac{1}{x^2} \, g(\ln x) \, y
\end{equation}
where
\begin{equation}\label{g}
g=\frac{f''-3f'}{f}
\end{equation}
The homogeneous solution with the correct asymptotic behavior at $x\rightarrow\infty$ is
\begin{equation}\label{y0}
y_0(x) = \left(1 + \frac{i}{x}\right)e^{ix}
\end{equation}
Therefore, using the Green's function method, Eq.~(\ref{yeq}) with the boundary condition Eq.~(\ref{bc}) can be written as the integral equation
\begin{equation}\label{inteq}
y(x) = y_0(x) + \frac{i}{2}\int_{x}^{\infty}du\,\frac{1}{u^2}\,g(\ln u)\,
y(u) \left[y_0^*(u)\,y_0(x)-y_0^*(x)\,y_0(u)\right]
\end{equation}
The power spectrum is
\begin{equation}\label{P}
\mathcal{P}_{\mathcal{R}_\mathrm{c}}
= \left(\frac{k}{2\pi}\right)^2 \lim_{x \rightarrow 0} \left|\frac{xy}{f}\right|^2
\end{equation}

\section{Series Formulae for the Spectrum}
\label{sf}

To solve Eq.~(\ref{inteq}) we expand $f(\ln x)$ and $g(\ln x)$ in power series
in $\ln(x/x_\star)$ where $x_\star$ is some convenient time around horizon crossing
\begin{equation}\label{fexp}
f(\ln x) = \sum_{n=0}^\infty \frac{f_n(x_\star)}{n!}
\lt[\ln\lt(\frac{x}{x_\star}\rt)\rt]^n
\end{equation}
\begin{equation}\label{gexp}
g(\ln x) = \sum_{n=0}^\infty \frac{g_{n+1}(x_\star)}{n!}
\lt[\ln\lt(\frac{x}{x_\star}\rt)\rt]^n
\end{equation}
This is similar to Stewart and Gong \cite{Jinook},
but I will make different assumptions about the magnitudes of the
$f_n$'s and $g_n$'s.
I assume $[f(\ln x)-f(\ln x_\star)]/f(\ln x_\star)$ and $g(\ln x)$ are small
around the time the mode $k$ leaves the horizon, \mbox{i.e.} for $\ln x$ of order one.
In terms of the $f_n$'s and $g_n$'s we have, for some small perturbation parameter $\xi$,
\begin{equation}\label{fxi}
\frac{f_n}{f_0} = \mathcal{O}(\xi)
\end{equation}
for $n\geq1$ and
\begin{equation}
g_n = \mathcal{O}(\xi)
\end{equation}
This is equivalent to Eqs.~(\ref{my1}) and~(\ref{my2}).
We want to solve Eq.~(\ref{inteq}) to order $\xi$,
\mbox{i.e.} to leading order in all the $g_n$'s.

\subsection{A particular solution}
\label{part}

To obtain the desired result I will use a mathematical trick.
Consider the special case
\begin{equation}\label{gsp}
g(\ln x) = \xi \lt(\frac{x}{x_\star}\rt)^\nu
\end{equation}
If $\nu=0$ or $\nu=1$ then Eq.~(\ref{yeq}) can be solved exactly.
See Refs.~\cite{foc} and~\cite{Martin} respectively.
However, we will be interested in the solution for arbitrary $\nu$ but only to leading order in $\xi$.
This is obtained by substituting Eqs.~(\ref{gsp}) and~(\ref{y0}) into the right hand side of Eq.~(\ref{inteq})
\begin{equation}
y(x) = y_0(x) + \frac{i}{2}\int_x^\infty du \,\frac{1}{u^2} \,
\xi x_\star^{-\nu} u^\nu \,y_0(u) \left[y_0^*(u)y_0(x)-y_0^*(x)y_0(u)\right]
+ \mathcal{O}\lt(\xi^2\rt)
\end{equation}
A little calculation gives
\begin{equation}
y(x) = y_0(x) + \frac{\xi x_\star^{-\nu}}{(3-\nu)(1-\nu)}
\left[ 2i x^{-1+\nu} e^{ix} - (1+\nu) \left(\frac{i}{2}\right)^\nu y_0^*(x)
\int_{-2ix}^\infty dv \,v^{-1+\nu} e^{-v} \right]
\end{equation}
The asymptotic behavior in the limit $x \rightarrow 0$ is
\begin{equation}
y(x) \stackrel{x \rightarrow 0}\longrightarrow
\frac{i}{x} \left\{1 + \frac{\xi x_\star^{-\nu}}{(3-\nu)\nu} \left[
\left(\frac{i}{2}\right)^\nu \frac{\Gamma(2+\nu)}{1-\nu} - x^\nu
\right] \right\}
\end{equation}
and so
\begin{equation}
\lim_{x\rightarrow 0} \left(\frac{xy}{f}\right) = \frac{i}{f}
\left\{ 1 + \frac{\xi x_\star^{-\nu}}{(3-\nu)\nu} \left[
\left(\frac{i}{2}\right)^\nu \frac{\Gamma(2+\nu)}{1-\nu} - x^\nu \right] \right\}
\end{equation}
This is constant to order $\xi$ and so we can evaluate it at any convenient time
around horizon crossing.
Evaluating at $x=x_\star$ and letting subscript $\star$ denote evaluation at that time,
we get
\begin{equation}
\lim_{x\rightarrow 0} \left(\frac{xy}{f}\right) = \frac{i}{f_\star}
\left\{ 1 + \frac{\xi}{(3-\nu)\nu} \left[
\left(\frac{i}{2x_\star}\right)^\nu \frac{\Gamma(2+\nu)}{1-\nu} - 1 \right] \right\}
\end{equation}
Eq.~(\ref{P}) gives
\begin{equation}\label{Asol}
\mathcal{P}_{\mathcal{R}_\mathrm{c}} = \left(\frac{k}{2\pi f_\star}\right)^2
\left[ 1 + 2 A_\star(\nu)\,\xi + \mathcal{O}\lt(\xi^2\rt) \right]
\end{equation}
where
\begin{equation}\label{Agen}
A_\star(\nu) \equiv \frac{1}{(3-\nu)\nu} \left[ (2x_\star)^{-\nu}
\cos\left(\frac{\pi\nu}{2}\right) \frac{\Gamma(2+\nu)}{1-\nu} - 1 \right]
\end{equation}
Note that $A_\star(\nu)$ is not singular at $\nu=0$ and can be expanded as
\begin{equation}\label{Aexp}
A_\star(\nu) = \sum_{n=0}^\infty a_n(x_\star)\,\nu^n
\end{equation}

\subsection{General solution}

Comparing
\begin{equation}
g(\ln x) = \xi \lt(\frac{x}{x_\star}\rt)^\nu = \xi e^{\nu\ln(x/x_\star)}
= \sum_{n=0}^\infty \frac{\xi\nu^n}{n!} \lt[\ln\lt(\frac{x}{x_\star}\rt)\rt]^n
\end{equation}
with the general expansion Eq.~(\ref{gexp})
\begin{equation}
g(\ln x) = \sum_{n=0}^\infty \frac{g_{n+1}}{n!}
\lt[\ln\lt(\frac{x}{x_\star}\rt)\rt]^n
\end{equation}
and noting that the solution of Section~\ref{part} was linear in $g$ throughout, we can equate the expansions {\em term by term\/} to get from Eqs.~(\ref{Asol}) and~(\ref{Aexp})
\begin{equation}
\mathcal{P}_{\mathcal{R}_\mathrm{c}}
= \left(\frac{k}{2\pi f_\star}\right)^2
\left[ 1 + 2 \sum_{n=0}^\infty a_n g_{n+1} + \mathcal{O}(\xi^2) \right]
\end{equation}
$A_\star(\nu)$ of Eq.~(\ref{Agen}) is the generating function for the $a_n$'s.
Defining $b_n \equiv a_{n-1}$ and
\begin{equation}
\sum_{n=0}^\infty b_n(x_\star)\,\nu^n \equiv B_\star(\nu) = \nu\,A_\star(\nu)
\end{equation}
we get
\begin{equation}\label{Pb}
\mathcal{P}_{\mathcal{R}_\mathrm{c}} = \left(\frac{k}{2\pi f_\star}\right)^2
\left[ 1 + 2 \sum_{n=1}^\infty b_n g_n + \mathcal{O}(\xi^2) \right]
\end{equation}

\subsection{Series formulae for the spectrum in terms of the slow-roll parameters}

Expressing $f$ and the $g_n$'s in terms of the slow-roll parameters
using Eqs.~(\ref{fsrV}), (\ref{g1}) and~(\ref{gn}),
we get from Eq.~(\ref{Pb})
\begin{equation}
\mathcal{P}_{\mathcal{R}_\mathrm{c}}
= \left(\frac{H_\star^4}{4\pi^2\dot{\phi}_\star^2}\right)
\left[ 1 + 2 \left( 6 b_1 - 1 \right) \epsilon_\star + 2 \sum_{n=1}^\infty (-1)^{n+1}
\left( 3 b_n - b_{n-1} \right) {\delta_n}_\star + \mathcal{O}(\xi^2) \right]
\end{equation}
where
\begin{equation}
\delta_n \equiv \frac{1}{H^n\dot{\phi}} \lt(\frac{d}{dt}\rt)^n \dot{\phi}
\end{equation}
Defining $d_0 \equiv 1$,
$d_n \equiv (-1)^n (3b_n-b_{n-1})$ for $n\geq1$ and
\begin{equation}
\sum_{n=0}^\infty d_n(x_\star)\,\nu^n \equiv D_\star(\nu)
= (3+\nu)\,B_\star(-\nu) + 1
\end{equation}
we get
\begin{eqnarray}\label{Pssr}
\mathcal{P}_{\mathcal{R}_\mathrm{c}}
& = & \left(\frac{H_\star}{2\pi}\right)^2 \left(\frac{H_\star}{\dot{\phi}_\star}\right)^2
\left[ 1 - (4d_1+2) \epsilon_\star - 2 \sum_{n=1}^\infty d_n {\delta_n}_\star
+ \mathcal{O}(\xi^2) \right]
\\
n_{\mathcal{R}_\mathrm{c}} - 1
& = & - 4 \epsilon_\star - 2 \sum_{n=0}^\infty d_n {\delta_{n+1}}_\star
+ \mathcal{O}(\xi^2)
\\
\frac{d n_{\mathcal{R}_\mathrm{c}}}{d\ln k}
& = & - 2 \sum_{n=0}^\infty d_n {\delta_{n+2}}_\star + \mathcal{O}(\xi^2)
\end{eqnarray}
Explicit formulae for the $d_n$'s are given in Appendix~\ref{Adn}.
Using Eqs.~(\ref{Hdot}), (\ref{phidot}), (\ref{edot}) and~(\ref{ddot})
we can rewrite this as
\begin{eqnarray}
\mathcal{P}_{\mathcal{R}_\mathrm{c}}
& = & \left(\frac{H_\star}{2\pi}\right)^2 \left(\frac{H_\star}{\dot{\phi}_\star}\right)^2
\left\{ 1 - (4d_1+2) \epsilon
- \lt[ D_\star\lt(\frac{1}{H}\frac{d}{dt}\rt) - 1 \rt] \ln\dot{\phi}^2
+ \mathcal{O}(\xi^2) \right\}_\star
\\
& = & \left(\frac{H_\star}{2\pi}\right)^2 \left(\frac{H_\star}{\dot{\phi}_\star}\right)^2
\left\{ 1 - (2d_1+2) \epsilon
- \lt[ D_\star\lt(\frac{1}{H}\frac{d}{dt}\rt) - 1 \rt] \ln\epsilon
+ \mathcal{O}(\xi^2) \right\}_\star
\\
& = & \left(\frac{H_\star}{2\pi}\right)^2 \left(\frac{H_\star}{\dot{\phi}_\star}\right)^2
\left\{ 1 - 2 \epsilon
+ \lt[ D_\star\lt(\frac{1}{H}\frac{d}{dt}\rt) - 1 \rt]
\ln\lt(\frac{H^4}{\dot{\phi}^2}\rt)
+ \mathcal{O}(\xi^2) \right\}_\star
\\
n_{\mathcal{R}_\mathrm{c}} - 1
& = & \left. - 4\epsilon - 2\,D_\star\lt(\frac{1}{H}\frac{d}{dt}\rt)\,\delta_1
+ \mathcal{O}(\xi^2) \right|_\star
\\
\frac{d n_{\mathcal{R}_\mathrm{c}}}{d\ln k}
& = & \left. - 2\,D_\star\lt(\frac{1}{H}\frac{d}{dt}\rt)\,\delta_2
+ \mathcal{O}(\xi^2) \right|_\star
\end{eqnarray}
where
\begin{equation}
D_\star(\nu) = (2x_\star)^\nu \cos\left(\frac{\pi\nu}{2}\right)
\frac{\Gamma(2-\nu)}{1+\nu}
\end{equation}
Standard slow-roll would correspond to setting $D_\star=1$, \mbox{i.e.} $d_n=0$
for $n\geq 1$, and neglecting the $-2\epsilon$ term in the formula for
$\mathcal{P}_{\mathcal{R}_\mathrm{c}}$.

\subsection{Series formulae in terms of the inflaton potential}
\label{sfp}

Expressing $f$ and the $g_n$'s in terms of the inflaton potential
using Eqs.~(\ref{fsrV}), (\ref{d1V}), (\ref{g1}) and~(\ref{gn}),
we get from Eq.~(\ref{Pb})
\begin{equation}
\mathcal{P}_{\mathcal{R}_\mathrm{c}}
= \frac{V_\star^3}{12\pi^2{V'_\star}^2}
\left\{ 1 + \left( 9 b_1 - \frac{1}{6} \right) \mathcal{U}_\star
- 2 \sum_{n=1}^\infty \left[ 3 b_n + \frac{1}{3^n} \right]
{\mathcal{V}_n}_\star + \mathcal{O}(\xi^2) \right\}
\end{equation}
where
\begin{eqnarray}
\mathcal{U} & \equiv & \left(\frac{V'}{V}\right)^2
\\
\mathcal{V}_n & \equiv & \left(\frac{V'}{V}\right)^{n-1} \frac{V^{(n+1)}}{V}
\end{eqnarray}
Defining $q_n \equiv 3 b_n + 1/3^n$ and
\begin{equation}
\sum_{n=0}^\infty q_n(x_\star)\,\nu^n \equiv Q_\star(\nu)
= 3 B_\star(\nu) + \frac{3}{3-\nu}
\end{equation}
we get
\begin{eqnarray}
\mathcal{P}_{\mathcal{R}_\mathrm{c}}
& = & \frac{V_\star^3}{12\pi^2{V'_\star}^2}
\left[ 1 + \left(3 q_1 - \frac{7}{6} \right) \mathcal{U}_\star
- 2 \sum_{n=1}^\infty q_n {\mathcal{V}_n}_\star + \mathcal{O}(\xi^2) \right]
\\ \label{nV}
n_{\mathcal{R}_\mathrm{c}} - 1 & = & - 3 \mathcal{U}_\star
+ 2 \sum_{n=0}^\infty q_n {\mathcal{V}_{n+1}}_\star + \mathcal{O}(\xi^2)
\\ \label{npV}
\frac{d n_{\mathcal{R}_\mathrm{c}}}{d\ln k}
& = & - 2 \sum_{n=0}^\infty q_n {\mathcal{V}_{n+2}}_\star + \mathcal{O}(\xi^2)
\end{eqnarray}
Explicit formulae for the $q_n$'s are given in Appendix~\ref{Aqn}.
We can rewrite this as
\begin{eqnarray}
\mathcal{P}_{\mathcal{R}_\mathrm{c}} & = & \frac{V_\star^3}{12\pi^2{V'_\star}^2}
\left\{ 1 + \lt(3q_1-\frac{7}{6}\rt) \mathcal{U}
- \lt[ Q_\star\lt(\frac{V'}{V}\frac{d}{d\phi}\rt) - 1 \rt] \ln{V'}^2
+ \mathcal{O}(\xi^2) \right\}_\star
\\ \label{sPVU}
& = & \frac{V_\star^3}{12\pi^2{V'_\star}^2}
\left\{ 1 + \lt(q_1-\frac{7}{6}\rt) \mathcal{U}
- \lt[ Q_\star\lt(\frac{V'}{V}\frac{d}{d\phi}\rt) - 1 \rt] \ln\mathcal{U}
+ \mathcal{O}(\xi^2) \right\}_\star
\\
& = & \frac{V_\star^3}{12\pi^2{V'_\star}^2} \left\{ 1 - \frac{7}{6} \mathcal{U}
+ \lt[ Q_\star\lt(\frac{V'}{V}\frac{d}{d\phi}\rt) - 1 \rt]
\ln\lt(\frac{V^3}{{V'}^2}\rt) + \mathcal{O}(\xi^2) \right\}_\star
\\
n_{\mathcal{R}_\mathrm{c}} - 1
& = & \left. - 3 \mathcal{U}
+ 2\,Q_\star\lt(\frac{V'}{V}\frac{d}{d\phi}\rt)\,\mathcal{V}_1
+ \mathcal{O}(\xi^2) \right|_\star
\\
\frac{d n_{\mathcal{R}_\mathrm{c}}}{d\ln k}
& = & \left. - 2\,Q_\star\lt(\frac{V'}{V}\frac{d}{d\phi}\rt)\,\mathcal{V}_2
+ \mathcal{O}(\xi^2) \right|_\star
\end{eqnarray}
where
\begin{equation}
Q_\star(\nu) = (2x_\star)^{-\nu} \cos\left(\frac{\pi\nu}{2}\right)
\frac{3\Gamma(2+\nu)}{(1-\nu)(3-\nu)}
\end{equation}
Standard slow-roll would correspond to setting $Q_\star=1$, \mbox{i.e.} $q_n=0$
for $n\geq 1$, and neglecting the $-7\mathcal{U}/6$ term in the formula for
$\mathcal{P}_{\mathcal{R}_\mathrm{c}}$.

\section{Integral Formulae for the Spectrum}

In this section I will derive integral formulae for the spectrum,
spectral index and running of the spectral index equivalent to
the series formulae of Section~\ref{sf}.

Eqs.~(\ref{P}) and~(\ref{inteq}) give
\begin{equation}
\mathcal{P}_{\mathcal{R}_\mathrm{c}}
= \left(\frac{k}{2\pi}\right)^2 \lim_{x \rightarrow 0} \frac{1}{f^2}
\lt| 1 + \frac{1}{2} x \int_x^\infty \frac{du}{u^2}\,g(\ln u)\,
y(u) \left[y_0^*(u)\,y_0(x)-y_0^*(x)\,y_0(u)\right] \rt|^2
\end{equation}
\begin{equation}
= \left(\frac{k}{2\pi}\right)^2 \lim_{x \rightarrow 0} \frac{1}{f^2}
\lt\{ 1 + \mathrm{Re} \int_x^\infty \frac{du}{u^2} g(\ln u)
y_0(u) x \left[ y_0^*(u) y_0(x) - y_0^*(x) y_0(u) \right]
+ \mathcal{O}\lt(g^2\rt) \rt\}
\end{equation}
Now from Eq.~(\ref{y0})
\begin{equation}
x\,y_0(x) = i + \frac{i}{2}x^2 - \frac{1}{3}x^3 + \mathcal{O}\lt(ix^4\rt)
\end{equation}
\begin{equation}
\lt|y_0(u)\rt|^2 = \frac{1}{u^2} \lt(1+u^2\rt)
\end{equation}
and
\begin{equation}
-\lt(y_0(u)\rt)^2 = \frac{1}{u^2} \lt\{ \lt[ \cos(2u) + 2u\sin(2u) - u^2\cos(2u) \rt]
+ i \lt[\sin(2u) - 2u\cos(2u) - u^2\sin(2u)\rt] \rt\}
\end{equation}
Therefore, assuming $g(\ln x)$ behaves like a function of $\ln x$
and not like a function of $x$,
\begin{equation}
\lim_{x\rightarrow0} \mathrm{Re} \lt[ x\,y_0(x)
\int_x^\infty  \frac{du}{u^2}\,g(\ln u) \lt|y_0(u)\rt|^2 \rt]
= - \frac{1}{3} \lim_{x\rightarrow0} x^3 \int_x^\infty \frac{du}{u^4}
\,g(\ln u) + \mathcal{O}\lt(x^2\rt)
\end{equation}
and
\begin{eqnarray}
\lefteqn{\lim_{x\rightarrow0} \mathrm{Re} \lt[ - x\,y_0^*(x)
\int_x^\infty \frac{du}{u^2}\,g(\ln u) \lt(y_0(u)\rt)^2 \rt]
= - \frac{1}{3} \lim_{x\rightarrow0} x^3 \int_x^\infty
\frac{du}{u^4}\,g(\ln u)}
\nn \\ && \mbox{}
+ \lim_{x\rightarrow0} \int_x^\infty \frac{du}{u}\,g(\ln u)
\lt[ \frac{\sin(2u)}{u^3} - \frac{2\cos(2u)}{u^2} - \frac{\sin(2u)}{u} \rt]
+ \mathcal{O}\lt(x^2\rt)
\end{eqnarray}
Therefore
\begin{equation}
\mathcal{P}_{\mathcal{R}_\mathrm{c}}
= \left(\frac{k}{2\pi}\right)^2 \lim_{x \rightarrow 0} \frac{1}{f^2}
\lt\{ 1 - \frac{2x^3}{3} \int_x^\infty \frac{du}{u^4}\,g(\ln u)
+ \frac{2}{3} \int_x^\infty \frac{du}{u}\,W(u)\,g(\ln u)
+ \mathcal{O}\lt(g^2\rt) \rt\}
\end{equation}
where
\begin{equation}
W(x) \equiv
\frac{3\sin(2x)}{2x^3} - \frac{3\cos(2x)}{x^2} - \frac{3\sin(2x)}{2x}
\end{equation}
Note that
\begin{equation}
\lim_{x \rightarrow 0} W(x) = 1 + \mathcal{O}\lt(x^2\rt)
\end{equation}
Letting subscript $\star$ denote evaluation at some convenient time around
horizon crossing,
\begin{equation}
\frac{1}{f^2} = \frac{1}{f_\star^2} \exp\lt[2\ln\lt(\frac{f_\star}{f}\rt)\rt]
= \frac{1}{f_\star^2} \lt[ 1 + 2 \int_x^{x_\star} \frac{du}{u} \frac{f'}{f}
+ \mathcal{O}\lt(g^2\rt) \rt]
\end{equation}
and
\begin{eqnarray}
- \frac{2}{3} x^3 \int_x^\infty \frac{du}{u^4}\,g(\ln u)
& = & 2 x^3 \int_x^\infty \frac{du}{u^4} \frac{f'}{f}
- \frac{2}{3} x^3 \int_x^\infty \frac{du}{u^4}\,\frac{f''}{f} \nn \\
& = & - \frac{2}{3} x^3 \lt[ \frac{1}{u^3} \frac{f'}{f} \rt]_x^\infty
+ \frac{2}{3} x^3 \int_x^\infty \frac{du}{u^4} \lt(\frac{f'}{f}\rt)'
- \frac{2}{3} x^3 \int_x^\infty \frac{du}{u^4}\,\frac{f''}{f} \nn \\
& = & \frac{2}{3} \frac{f'_\star}{f_\star}
- \frac{2}{3} \int_x^{x_\star} \frac{du}{u} \lt(\frac{f'}{f}\rt)'
+ \mathcal{O}\lt(g^2\rt) \nn \\
& = & \frac{2}{3} \frac{f'_\star}{f_\star}
- \frac{2}{3} \int_x^{x_\star} \frac{du}{u} \frac{f''}{f}
+ \mathcal{O}\lt(g^2\rt)
\end{eqnarray}
Therefore
\begin{equation}\label{iPfg}
\mathcal{P}_{\mathcal{R}_\mathrm{c}}
= \left(\frac{k}{2\pi}\right)^2 \frac{1}{f_\star^2}
\lt\{ 1 + \frac{2}{3} \frac{f'_\star}{f_\star}
+ \frac{2}{3} \int_0^\infty \frac{du}{u} \lt[W(u)-\theta(x_\star-u)\rt] g(\ln u)
+ \mathcal{O}\lt(g^2\rt) \rt\}
\end{equation}
where $\theta(x) = 0$ for $x<0$ and $\theta(x) = 1$ for $x>0$.

\subsection{Integral formulae in terms of the slow-roll parameters}

Now
\begin{equation}
g = \frac{f''}{f} - 3\frac{f'}{f}
= \lt(\frac{f'}{f}\rt)' - 3\frac{f'}{f} + \mathcal{O}\lt(g^2\rt)
\end{equation}
and
\begin{equation}
\int_0^\infty \frac{du}{u} \lt[ W(u) - \theta\lt(x_\star-u\rt) \rt]
\lt[\frac{f'(\ln u)}{f(\ln u)}\rt]'
= - \frac{f'_\star}{f_\star} - \int_0^\infty du\,W'(u)\,\frac{f'(\ln u)}{f(\ln u)}
\end{equation}
Therefore, substituting into Eq.~(\ref{iPfg}), we get
\begin{equation}
\mathcal{P}_{\mathcal{R}_\mathrm{c}}
= \left(\frac{k}{2\pi}\right)^2 \frac{1}{f_\star^2}
\lt\{ 1 - 2 \int_0^\infty \frac{du}{u} \lt[ \omega(u) - \theta\lt(x_\star-u\rt) \rt]
\frac{f'(\ln u)}{f(\ln u)} \rt\}
\end{equation}
where
\begin{equation}
\omega(x) \equiv W(x) + \frac{x}{3}\,W'(x)
\end{equation}
Note that
\begin{equation}
\lim_{x \rightarrow 0} \omega(x) = 1 + \mathcal{O}\lt(x^2\rt)
\end{equation}
Substituting Eqs.~(\ref{fsrV}), (\ref{x}) and~(\ref{fp}) gives
\begin{equation}
\mathcal{P}_{\mathcal{R}_\mathrm{c}}
= \frac{H_\star^4}{4\pi^2\dot{\phi}_\star^2}
\lt\{ 1 - 2 \epsilon_\star + 2 \int_0^\infty \frac{du}{u}
\lt[\omega(u)-\theta\lt(\frac{k}{a_\star H_\star}-u\rt)\rt] (2\epsilon+\delta_1)|_{aH=k/u}
+ \mathcal{O}\lt(\xi^2\rt) \rt\}
\end{equation}
Now, Eq.~(\ref{econst}) shows that $\epsilon$ is approximately constant, and
\begin{equation}
\int_0^\infty \frac{du}{u}
\lt[ \omega(u) - \theta\lt(\frac{k}{a_\star H_\star}-u\rt) \rt]
= \alpha - \ln\lt(\frac{k}{a_\star H_\star}\rt)
= - d_1
\end{equation}
where $\alpha \equiv 2-\ln 2-\gamma \simeq 0.729637$
and $\gamma$ is the Euler-Mascheroni constant.
Therefore to order $\xi$
\begin{eqnarray}\label{Pisr}
\mathcal{P}_{\mathcal{R}_\mathrm{c}}
& = & \frac{H_\star^4}{4\pi^2\dot{\phi}_\star^2} \lt\{ 1 - \lt(4d_1+2\rt) \epsilon_\star
+ 2 \int_0^\infty \frac{du}{u}
\lt[\omega(u)-\theta\lt(\frac{k}{a_\star H_\star}-u\rt)\rt]
\delta_1|_{aH=\frac{k}{u}} \rt\}
\\
n_{\mathcal{R}_\mathrm{c}} - 1
& = & - 4 \epsilon_\star - 2 {\delta_1}_\star + 2 \int_0^\infty \frac{du}{u}
\lt[\omega(u)-\theta\lt(\frac{k}{a_\star H_\star}-u\rt)\rt]
\lt.\delta_2\rt|_{aH=\frac{k}{u}}
\\
\frac{d n_{\mathcal{R}_\mathrm{c}}}{d\ln k}
& = & - 2 {\delta_2}_\star + 2 \int_0^\infty \frac{du}{u}
\lt[\omega(u)-\theta\lt(\frac{k}{a_\star H_\star}-u\rt)\rt]
\lt.\delta_3\rt|_{aH=\frac{k}{u}}
\end{eqnarray}
where
\begin{equation}
\omega(x) = \frac{\sin(2x)}{x} - \cos(2x)
\end{equation}

\subsection{Integral formulae in terms of the inflaton potential}

Substituting Eqs.~(\ref{fsrV}), (\ref{fp}), (\ref{x}) and~(\ref{gsrV}) into Eq.~(\ref{iPfg}) gives
\begin{equation}
\mathcal{P}_{\mathcal{R}_\mathrm{c}} = \frac{V_\star^3}{12\pi^2{V'_\star}^2}
\lt\{ 1 - \frac{7}{6} \mathcal{U}_\star
- 2 \int_0^\infty \frac{du}{u} \lt[W(u)-\theta\lt(\frac{k}{a_\star H_\star}-u\rt)\rt]
\lt.\lt( \mathcal{V}_1 - \frac{3}{2} \mathcal{U} \rt)\rt|_{aH=\frac{k}{u}}
+ \mathcal{O}\lt(\xi^2\rt) \rt\}
\end{equation}
Now, $\mathcal{U}$ is approximately constant and
\begin{equation}
\int_0^\infty \frac{du}{u} \lt[ W(u) - \theta\lt(\frac{k}{a_\star H_\star}-u\rt) \rt]
= \alpha + \frac{1}{3} - \ln\lt(\frac{k}{a_\star H_\star}\rt)
= q_1
\end{equation}
Therefore to order $\xi$
\begin{equation}\label{iPV1}
\mathcal{P}_{\mathcal{R}_\mathrm{c}} = \frac{V_\star^3}{12\pi^2{V'_\star}^2}
\lt\{ 1 + \lt(3q_1-\frac{7}{6}\rt) \mathcal{U}_\star
- 2 \int_0^\infty \frac{du}{u} \lt[W(u)-\theta\lt(\frac{k}{a_\star H_\star}-u\rt)\rt]
\lt.\mathcal{V}_1\rt|_{aH=\frac{k}{u}} \rt\}
\end{equation}
\begin{eqnarray}\label{inV1}
n_{\mathcal{R}_\mathrm{c}} - 1 & = & - 3 \mathcal{U}_\star + 2 {\mathcal{V}_1}_\star
+ 2 \int_0^\infty \frac{du}{u} \lt[W(u)-\theta\lt(\frac{k}{a_\star H_\star}-u\rt)\rt]
\lt.\mathcal{V}_2\rt|_{aH=\frac{k}{u}}
\\ \label{inpV1}
\frac{d n_{\mathcal{R}_\mathrm{c}}}{d\ln k} & = & - 2 {\mathcal{V}_2}_\star
- 2 \int_0^\infty \frac{du}{u} \lt[W(u)-\theta\lt(\frac{k}{a_\star H_\star}-u\rt)\rt]
\lt.\mathcal{V}_3\rt|_{aH=\frac{k}{u}}
\end{eqnarray}
Integrating by parts gives to order $\xi$
\begin{eqnarray}
\mathcal{P}_{\mathcal{R}_\mathrm{c}}
& = & \frac{V_\star^3}{12\pi^2{V'_\star}^2}
\lt\{ 1 + \lt(3q_1-\frac{7}{6}\rt) \mathcal{U}_\star + 2 \int_0^\infty du\,W'(u)
\lt.\ln\lt(\frac{V'}{V'_\star}\rt)\rt|_{aH=\frac{k}{u}} \rt\}
\\ \label{iPV2}
& = & \frac{V_\star^3}{12\pi^2{V'_\star}^2}
\lt\{ 1 + \lt(q_1-\frac{7}{6}\rt) \mathcal{U}_\star + \int_0^\infty du\,W'(u)
\lt.\ln\lt(\frac{\mathcal{U}}{\mathcal{U}_\star}\rt)\rt|_{aH=\frac{k}{u}} \rt\}
\\
& = & \frac{V_\star^3}{12\pi^2{V'_\star}^2}
\lt\{ 1 - \frac{7}{6} \mathcal{U}_\star - \int_0^\infty du\,W'(u)
\lt[ \lt.\ln\lt(\frac{V^3}{{V'}^2}\rt)\rt|_{aH=\frac{k}{u}}
- \ln\lt(\frac{V_\star^3}{{V'_\star}^2}\rt) \rt] \rt\}
\\
n_{\mathcal{R}_\mathrm{c}} - 1 & = & - 3 \mathcal{U}_\star
- 2 \int_0^\infty du\,W'(u) \lt.\mathcal{V}_1\rt|_{aH=\frac{k}{u}}
\\
\frac{d n_{\mathcal{R}_\mathrm{c}}}{d\ln k}
& = & 2 \int_0^\infty du\,W'(u) \lt.\mathcal{V}_2\rt|_{aH=\frac{k}{u}}
\end{eqnarray}
where
\begin{equation}
W(x) =
\frac{3\sin(2x)}{2x^3} - \frac{3\cos(2x)}{x^2} - \frac{3\sin(2x)}{2x}
\end{equation}
and
\begin{eqnarray}
- x\,W'(x) = \frac{9\sin(2x)}{2x^3} - \frac{9\cos(2x)}{x^2}
- \frac{15\sin(2x)}{2x} + 3\cos(2x)
\end{eqnarray}
Note that
\begin{equation}
\lim_{x \rightarrow 0} \lt[- x\,W'(x)\rt]
= - \frac{4}{5} x^2 + \mathcal{O}\lt(x^4\rt)
\end{equation}
and
\begin{equation}
- \int_0^\infty dx\,W'(x) = 1
\end{equation}

\begin{figure}[h]\label{window}
\begin{center}
\epsfig{file=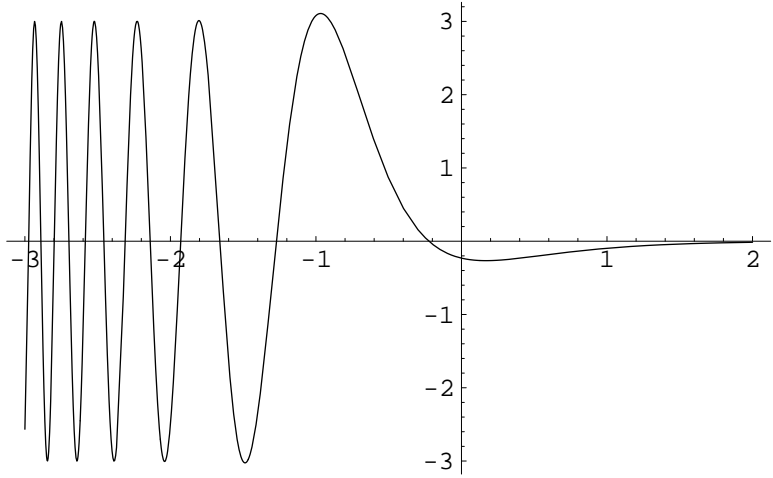}
\end{center}
\caption{The window function $-\frac{k}{aH}\,W'\lt(\frac{k}{aH}\rt)$
as a function of $\ln\lt(\frac{aH}{k}\rt)$.}
\end{figure}

\section{Optimized Standard Slow-Roll}
\label{op}

If we choose
\begin{equation}
a_\star H_\star = e^{-\alpha-\frac{1}{3}} k \simeq e^{-1.063} k \simeq 0.345\,k
\end{equation}
we can set
\begin{equation}
q_1 = \int_0^\infty \frac{du}{u}
\lt[ W(u) - \theta\lt(\frac{k}{a_\star H_\star}-u\rt) \rt] = 0
\end{equation}
Neglecting the integrals in Eqs.~(\ref{iPV1}), (\ref{inV1}) and~(\ref{inpV1})
then gives
\begin{eqnarray}
\mathcal{P}_{\mathcal{R}_\mathrm{c}} & = & \left.
\frac{V^3}{12\pi^2{V'}^2} \lt( 1 - \frac{7}{6} \mathcal{U} \rt)
\right|_{aH=k/\exp\lt(\alpha+\frac{1}{3}\rt)}
\\
n_{\mathcal{R}_\mathrm{c}} - 1
& = & \left. - 3 \mathcal{U} + 2 {\mathcal{V}_1}
\right|_{aH=k/\exp\lt(\alpha+\frac{1}{3}\rt)}
\\
\frac{d n_{\mathcal{R}_\mathrm{c}}}{d\ln k}
& = & \left. - 2 {\mathcal{V}_2} \right|_{aH=k/\exp\lt(\alpha+\frac{1}{3}\rt)}
\end{eqnarray}
which I will call ``optimized standard slow-roll''.
See Figure~\ref{comparing}.

\section{Examples}

Consider the class of potentials
\footnote{This form is chosen for illustrative convenience.
Similar but more realistic potentials include those of the form
$V = V_0 [1 + A_1\,f_1(\phi/\Lambda_1) + A_2\,f_2(\phi/\Lambda_2)]$
with $A_2 \ll A_1 \ll 1$ and $\Lambda_1 \gg \Lambda_2$.}
\begin{equation}
V = V_0 e^{\lambda\phi} \left[ 1 + A \, f(\nu\phi) \right]
\end{equation}
where we have in mind $|\lambda| \ll 1$, $|A| \ll 1$, $|\nu| \gg 1$,
and $f(x)$ is some smooth function.
Now
\begin{equation}
\frac{V'}{V} = \lambda + \frac{A\nu f'}{1+Af}
\end{equation}
Therefore, assuming
\begin{eqnarray}
\lambda^2 & = & \mathcal{O}\lt(\xi\rt)
\\
\frac{A\nu}{\lambda} f' & = & \mathcal{O}\lt(\xi\rt)
\\
A f & = & \mathcal{O}\lt(\xi^2\rt)
\end{eqnarray}
we have
\begin{equation}
\frac{V^3}{{V'}^2} = \frac{V_0 e^{\lambda\phi}}{\lambda^2}
\lt[ 1 - \frac{2A\nu}{\lambda} f' + \mathcal{O}\lt(\xi^2\rt) \rt]
\end{equation}
and
\begin{equation}
\mathcal{U} = \lambda^2 \lt[ 1 + \frac{2A\nu}{\lambda} f' + \mathcal{O}\lt(\xi^2\rt) \rt]
\end{equation}
Substituting into Eq.~(\ref{sPVU}) gives
\begin{equation}
\mathcal{P}_{\mathcal{R}_\mathrm{c}}
= \frac{V_0 e^{\lambda\phi_\star}}{12\pi^2\lambda^2}
\left[ 1 + \left(q_1-\frac{7}{6}\right) \lambda^2
- \frac{2A\nu}{\lambda}\,Q_\star\lt(\lambda\frac{d}{d\phi}\rt)\,f'(\nu\phi)
+ \mathcal{O}(\xi^2) \right]_\star
\end{equation}
while substituting into Eq.~(\ref{iPV2}) gives
\begin{equation}
\mathcal{P}_{\mathcal{R}_\mathrm{c}}
= \frac{V_0 e^{\lambda\phi_\star}}{12\pi^2\lambda^2}
\lt[ 1 + \lt( q_1 - \frac{7}{6} \rt) \lambda^2
+ \frac{2A\nu}{\lambda} \int_0^\infty du\,W'(u)
\lt.f'(\nu\phi)\rt|_{aH=\frac{k}{u}}
+ \mathcal{O}\lt(\xi^2\rt) \rt]
\end{equation}
Now
\begin{equation}
\frac{d\phi}{d\ln(aH)} = \frac{\dot\phi}{(1-\epsilon)H}
= - \frac{V'}{V} \lt[ 1 + \mathcal{O}(\xi) \rt]
= - \lambda \lt[ 1 + \mathcal{O}(\xi) \rt]
\end{equation}
Therefore
\begin{equation}
\phi \simeq \lambda \ln\lt(\frac{a_0H_0}{aH}\rt) + \phi_0
\end{equation}
Defining
\begin{equation}
\mathcal{N} \equiv \ln\lt(\frac{a_0H_0}{k}\rt)
\end{equation}
and setting $\phi_0=0$, we have
\begin{eqnarray}
\phi & \simeq & \lambda \mathcal{N} + \lambda \ln\lt(\frac{k}{aH}\rt)
\\
\phi_\star & \simeq & \lambda \mathcal{N} + \lambda \ln x_\star
\\
\phi|_{aH=\frac{k}{u}} & \simeq & \lambda \mathcal{N} + \lambda \ln u
\end{eqnarray}
Therefore, to order $\xi$,
\begin{eqnarray}\label{sPex}
\mathcal{P}_{\mathcal{R}_\mathrm{c}} & = &
\frac{V_0 e^{\lambda^2\mathcal{N}}}{12\pi^2\lambda^2}
\left[ 1 + \left(\alpha - \frac{5}{6} \right) \lambda^2
- \frac{2A\nu}{\lambda}\,Q_\star\lt(\frac{d}{d\mathcal{N}}\rt)
f'(\lambda\nu\mathcal{N}+\lambda\nu\ln x_\star) \right]
\\
n_{\mathcal{R}_\mathrm{c}} - 1 & = & - \lambda^2
+ \frac{2A\nu}{\lambda}\,\lambda\nu\,Q_\star\lt(\frac{d}{d\mathcal{N}}\rt)
f''(\lambda\nu\mathcal{N}+\lambda\nu\ln x_\star)
\\
\frac{d n_{\mathcal{R}_\mathrm{c}}}{d\ln k} & = &
- \frac{2A\nu}{\lambda}\,(\lambda\nu)^2\,Q_\star\lt(\frac{d}{d\mathcal{N}}\rt)
f'''(\lambda\nu\mathcal{N}+\lambda\nu\ln x_\star)
\end{eqnarray}
where we will usually choose to set $x_\star=1$.
The equivalent integral formulae are
\begin{eqnarray}\label{iPex}
\mathcal{P}_{\mathcal{R}_\mathrm{c}}
& = & \frac{V_0 e^{\lambda^2\mathcal{N}}}{12\pi^2\lambda^2}
\lt[ 1 + \lt( \alpha - \frac{5}{6} \rt) \lambda^2
+ \frac{2A\nu}{\lambda} \int_0^\infty du\,W'(u)\,
f'(\lambda\nu\mathcal{N} + \lambda\nu\ln u) \rt]
\\
n_{\mathcal{R}_\mathrm{c}} - 1
& = & - \lambda^2 - \frac{2A\nu}{\lambda}\,\lambda\nu
\int_0^\infty du\,W'(u)\,f''(\lambda\nu\mathcal{N} + \lambda\nu\ln u)
\\
\frac{d n_{\mathcal{R}_\mathrm{c}}}{d\ln k}
& = & \frac{2A\nu}{\lambda}\,(\lambda\nu)^2
\int_0^\infty du\,W'(u)\,f'''(\lambda\nu\mathcal{N} + \lambda\nu\ln u)
\end{eqnarray}

For comparison, standard slow-roll gives
\begin{eqnarray}\label{ssrPex}
\mathcal{P}_{\mathcal{R}_\mathrm{c}}
& = & \frac{V_0 e^{\lambda^2\mathcal{N}}}{12\pi^2\lambda^2}
\lt[ 1 - \frac{2A\nu}{\lambda}\,f'(\lambda\nu\mathcal{N}) \rt]
\\
n_{\mathcal{R}_\mathrm{c}} - 1 & = & - \lambda^2
+ \frac{2A\nu}{\lambda}\,\lambda\nu\,f''(\lambda\nu\mathcal{N})
\\
\frac{d n_{\mathcal{R}_\mathrm{c}}}{d\ln k}
& = & - \frac{2A\nu}{\lambda}\,(\lambda\nu)^2\,f'''(\lambda\nu\mathcal{N})
\end{eqnarray}
while optimized standard slow-roll, Section~\ref{op}, gives
\begin{eqnarray}\label{osrPex}
\mathcal{P}_{\mathcal{R}_\mathrm{c}}
& = & \frac{V_0 e^{\lambda^2\mathcal{N}_\star}}{12\pi^2\lambda^2}
\lt[ 1 - \frac{7}{6} \lambda^2
- \frac{2A\nu}{\lambda}\,f'(\lambda\nu\mathcal{N}_\star) \rt]
\\
n_{\mathcal{R}_\mathrm{c}} - 1 & = & - \lambda^2
+ \frac{2A\nu}{\lambda}\,\lambda\nu\,f''(\lambda\nu\mathcal{N}_\star)
\\
\frac{d n_{\mathcal{R}_\mathrm{c}}}{d\ln k}
& = & - \frac{2A\nu}{\lambda}\,(\lambda\nu)^2\,f'''(\lambda\nu\mathcal{N}_\star)
\end{eqnarray}
where
\begin{equation}
\mathcal{N}_\star \equiv \mathcal{N} + \alpha + \frac{1}{3}
\end{equation}

We now consider some specific choices of $f$.

\subsection{Power}

For
\begin{equation}
f(x) = x^p
\end{equation}
we have
\begin{equation}
f^{(n)}(x) = \frac{p!}{(p-n)!} x^{p-n}
\end{equation}
Therefore Eq.~(\ref{sPex}) gives
\begin{equation}
\mathcal{P}_{\mathcal{R}_\mathrm{c}}
= \frac{V_0 e^{\lambda^2\mathcal{N}}}{12\pi^2\lambda^2}
\left[ 1 + \left(\alpha - \frac{5}{6} \right) \lambda^2
- 2 \lt(\frac{A\nu}{\lambda}\rt) p! (\lambda\nu)^{p-1}
\sum_{l=0}^{p-1} \frac{q_{p-1-l}}{l!} \mathcal{N}^l
\right]
\end{equation}
\begin{equation}
n_{\mathcal{R}_\mathrm{c}} - 1 = - \lambda^2
+ 2 \lt(\frac{A\nu}{\lambda}\rt) p! (\lambda\nu)^{p-1}
\sum_{l=0}^{p-2} \frac{q_{p-2-l}}{l!} \mathcal{N}^l
\end{equation}
and
\begin{equation}
\frac{d n_{\mathcal{R}_\mathrm{c}}}{d\ln k} =
- 2 \lt(\frac{A\nu}{\lambda}\rt) p! (\lambda\nu)^{p-1}
\sum_{l=0}^{p-3} \frac{q_{p-3-l}}{l!} \mathcal{N}^l
\end{equation}
where the $q_n$'s should be evaluated at $x_\star=1$.

\subsection{Power series}

For
\begin{equation}
f(x) = \sum_{p=0}^\infty \frac{f_p}{p!} x^p
\end{equation}
we have
\begin{equation}
f^{(n)}(x) = \sum_{p=0}^\infty \frac{f_{p+n}}{p!} x^p
\end{equation}
Therefore Eq.~(\ref{sPex}) gives
\begin{equation}
\mathcal{P}_{\mathcal{R}_\mathrm{c}}
= \frac{V_0 e^{\lambda^2\mathcal{N}}}{12\pi^2\lambda^2}
\left[ 1 + \left(\alpha - \frac{5}{6} \right) \lambda^2
- 2 \lt(\frac{A\nu}{\lambda}\rt)
\sum_{n=0}^\infty \sum_{p=0}^\infty f_{p+n+1}\,q_n (\lambda\nu)^n
\frac{(\lambda\nu\mathcal{N})^p}{p!} \right]
\end{equation}
\begin{equation}
n_{\mathcal{R}_\mathrm{c}} - 1 = - \lambda^2
+ 2 \lt(\frac{A\nu}{\lambda}\rt) \lambda\nu
\sum_{n=0}^\infty \sum_{p=0}^\infty f_{p+n+2}\,q_n (\lambda\nu)^n
\frac{(\lambda\nu\mathcal{N})^p}{p!}
\end{equation}
and
\begin{equation}
\frac{d n_{\mathcal{R}_\mathrm{c}}}{d\ln k} =
- 2 \lt(\frac{A\nu}{\lambda}\rt) (\lambda\nu)^2
\sum_{n=0}^\infty \sum_{p=0}^\infty f_{p+n+3}\,q_n (\lambda\nu)^n
\frac{(\lambda\nu\mathcal{N})^p}{p!}
\end{equation}
where the $q_n$'s should be evaluated at $x_\star=1$.

\subsection{Exponential}

For
\begin{equation}
f(x) = e^x
\end{equation}
Eq.~(\ref{sPex}) gives
\begin{equation}
\mathcal{P}_{\mathcal{R}_\mathrm{c}}
= \frac{V_0 e^{\lambda^2\mathcal{N}}}{12\pi^2\lambda^2}
\left[ 1 + \left(\alpha - \frac{5}{6} \right) \lambda^2
- 2 \lt(\frac{A\nu}{\lambda}\rt)
\, Q(\lambda\nu) \, e^{\lambda\nu\mathcal{N}} \right]
\end{equation}
and
\begin{equation}
n_{\mathcal{R}_\mathrm{c}} - 1 = - \lambda^2
+ 2 \lt(\frac{A\nu}{\lambda}\rt) \lambda\nu
\, Q(\lambda\nu) \, e^{\lambda\nu\mathcal{N}}
\end{equation}
where
\begin{equation}
Q(\nu) \equiv \lt.Q_\star(\nu)\rt|_{x_\star=1}
\end{equation}

\subsection{Sine}

For
\begin{equation}
f(x) = \sin x = \mathrm{Im}\, e^{ix}
\end{equation}
Eq.~(\ref{sPex}) gives
\begin{equation}
\mathcal{P}_{\mathcal{R}_\mathrm{c}}
= \frac{V_0 e^{\lambda^2\mathcal{N}}}{12\pi^2\lambda^2}
\lt\{ 1 + \left(\alpha - \frac{5}{6} \right) \lambda^2
- 2 \lt(\frac{A\nu}{\lambda}\rt)
\mathrm{Re}\lt[Q(i\lambda\nu)\,e^{i\lambda\nu\mathcal{N}}\rt] \rt\}
\end{equation}
and
\begin{equation}
n_{\mathcal{R}_\mathrm{c}} - 1 = - \lambda^2
- 2 \lt(\frac{A\nu}{\lambda}\rt) \lambda\nu
\, \mathrm{Im}\lt[Q(i\lambda\nu)\,e^{i\lambda\nu\mathcal{N}}\rt]
\end{equation}

\subsection{Fourier transform}

For
\begin{equation}
f(x) = \int_{-\infty}^\infty \tilde{f}(p)\,e^{ipx}\,dp
\end{equation}
we have
\begin{equation}
f^{(n)}(x) = \int_{-\infty}^\infty (ip)^n \tilde{f}(p)\,e^{ipx}\,dp
\end{equation}
Therefore Eq.~(\ref{sPex}) gives
\begin{equation}\label{Fourier}
\mathcal{P}_{\mathcal{R}_\mathrm{c}}
= \frac{V_0 e^{\lambda^2\mathcal{N}}}{12\pi^2\lambda^2}
\lt\{ 1 + \left(\alpha - \frac{5}{6} \right) \lambda^2
- 2 \lt(\frac{A\nu}{\lambda}\rt)
\int_{-\infty}^\infty ip \,\tilde{f}(p)\,
Q(ip\lambda\nu)\,e^{ip\lambda\nu\mathcal{N}} dp \rt\}
\end{equation}
\begin{equation}
n_{\mathcal{R}_\mathrm{c}} - 1 = - \lambda^2
+ 2 \lt(\frac{A\nu}{\lambda}\rt) \lambda\nu
\int_{-\infty}^\infty (ip)^2 \tilde{f}(p)\,
Q(ip\lambda\nu)\,e^{ip\lambda\nu\mathcal{N}} dp
\end{equation}
and
\begin{equation}
\frac{d n_{\mathcal{R}_\mathrm{c}}}{d\ln k} =
- 2 \lt(\frac{A\nu}{\lambda}\rt) (\lambda\nu)^2
\int_{-\infty}^\infty (ip)^3 \tilde{f}(p)\,
Q(ip\lambda\nu)\,e^{ip\lambda\nu\mathcal{N}} dp
\end{equation}

\subsection{Step}
\label{step}

\begin{figure}[p]
\begin{center}
\epsfig{file=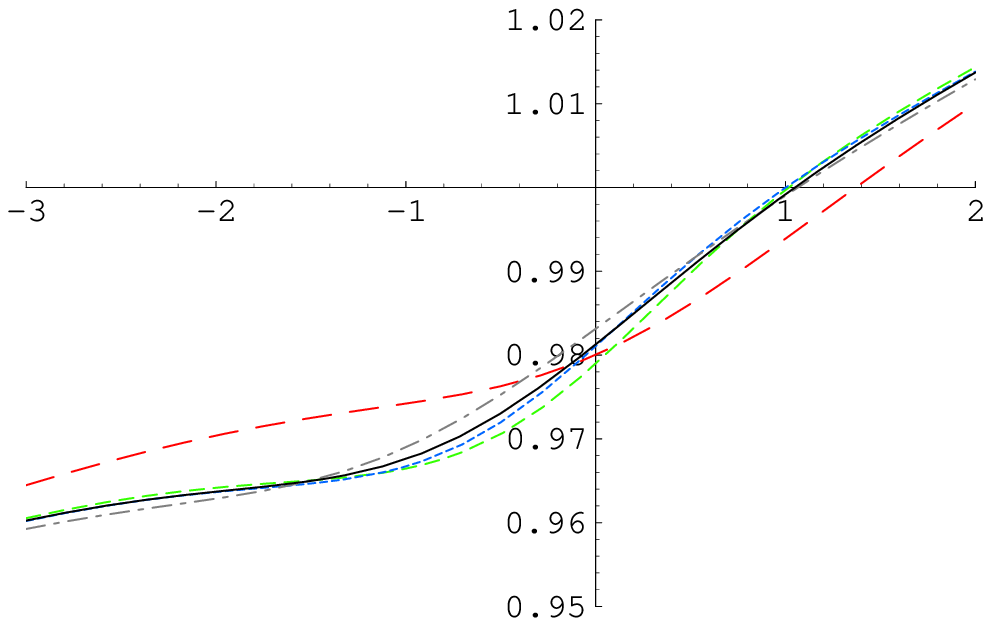,width=8cm}
\epsfig{file=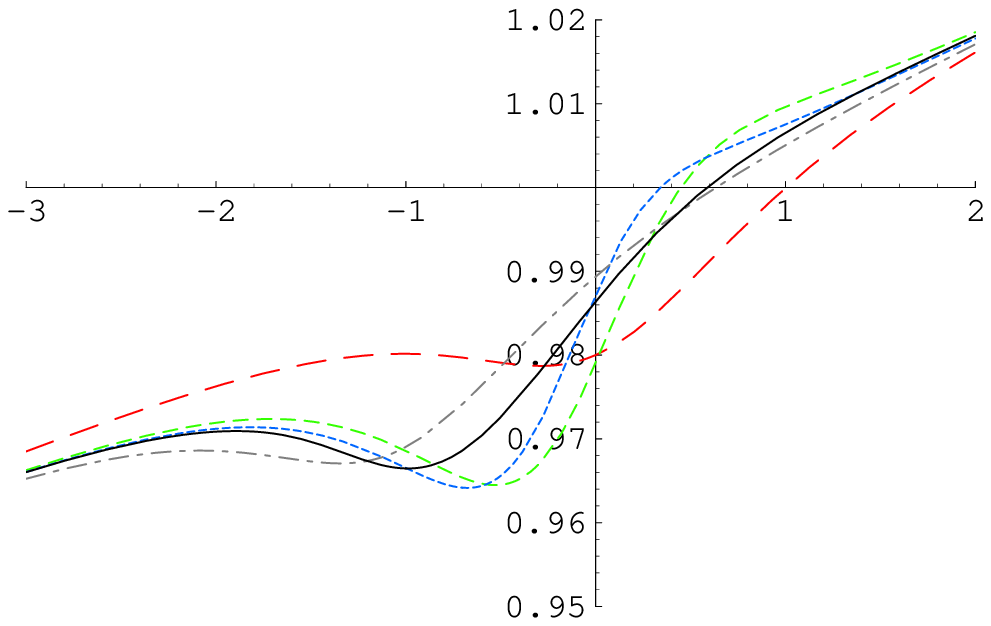,width=8cm}
\epsfig{file=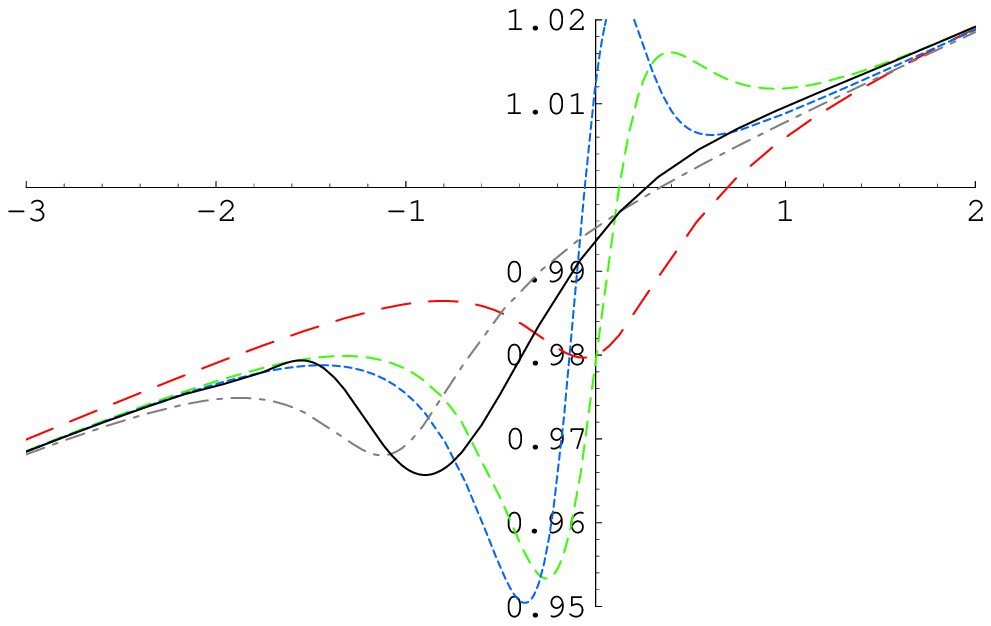,width=8cm}
\epsfig{file=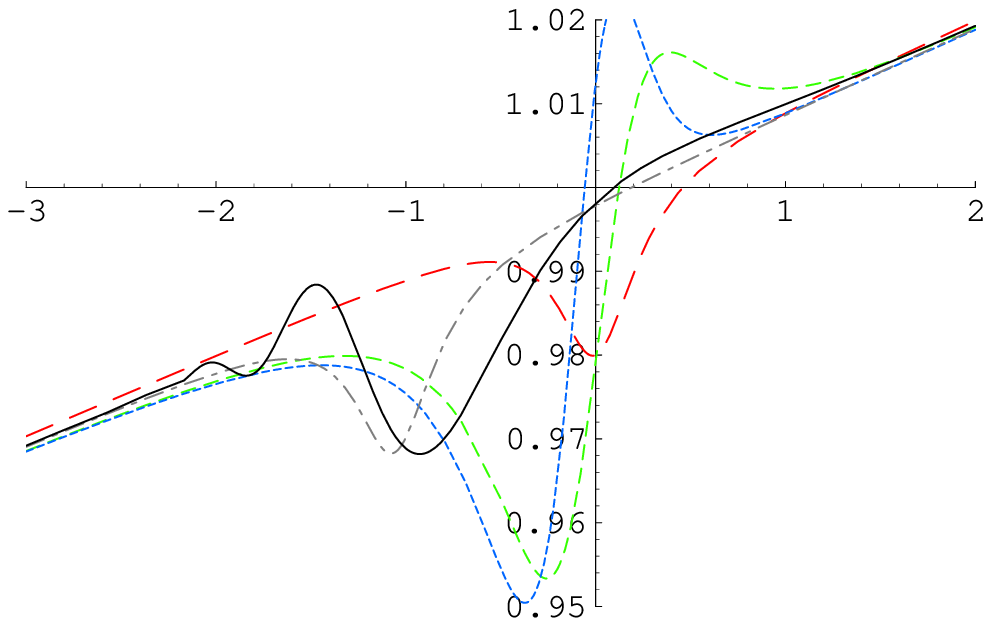,width=8cm}
\end{center}
\caption{\label{comparing}
Spectrum for the arctangent step of Eq.~(\ref{astep}) using
standard slow-roll (long dash),
standard slow-roll with 1st order corrections \cite{foc} (dash) and
standard slow-roll with 2nd order corrections \cite{Jinook} (short dash),
all evaluated at $aH=k$, and
optimized standard slow-roll (dot dash)
and general slow-roll (solid).
The parameters are $\lambda=0.1$, $A=\lambda^3/\nu$ and
$\lambda\nu=\{0.5,1,2,4\}$, top left to bottom right.
Standard slow-roll fails to converge for $\lambda\nu\gtrsim 2$,
while optimized standard slow-roll completely misses the ringing
that occurs for $\lambda\nu\gtrsim 4$.}
\end{figure}

\begin{figure}[p]
\begin{center}
\epsfig{file=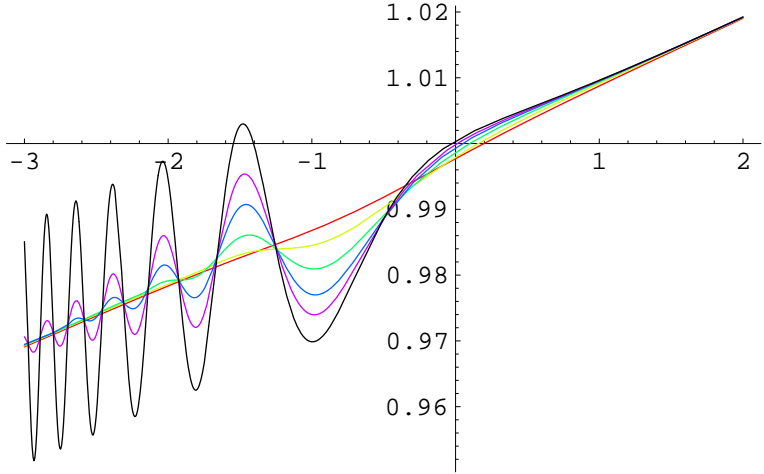}
\end{center}
\caption{\label{steepening}
Spectrum for steepening arctangent steps,
with $\lambda=0.1$, $A=10^{-5}$ and $\nu=\{10,20,40,80,160\}$,
enclosed by the sharp step limit of Eq.~(\ref{sharplimit}).}
\end{figure}

For an arctangent step
\begin{equation}\label{astep}
f(x) = \arctan x
\end{equation}
we have
\begin{equation}
\tilde{f}(p) = \frac{1}{2ip} e^{-|p|}
\end{equation}
Therefore Eq.~(\ref{Fourier}) gives
\begin{equation}
\mathcal{P}_{\mathcal{R}_\mathrm{c}}
= \frac{V_0 e^{\lambda^2\mathcal{N}}}{12\pi^2\lambda^2}
\lt\{ 1 + \left(\alpha - \frac{5}{6} \right) \lambda^2
- 2 \lt(\frac{A\nu}{\lambda}\rt)
\int_0^\infty e^{-p}\,\mathrm{Re}\lt[
Q(ip\lambda\nu)\,e^{ip\lambda\nu\mathcal{N}}\rt] dp \rt\}
\end{equation}
Alternatively Eq.~(\ref{iPex}) gives
\begin{equation}
\mathcal{P}_{\mathcal{R}_\mathrm{c}}
= \frac{V_0 e^{\lambda^2\mathcal{N}}}{12\pi^2\lambda^2} \lt\{ 1
+ \lt( \alpha - \frac{5}{6} \rt) \lambda^2
+ 2 \lt(\frac{A\nu}{\lambda}\rt) \int_0^\infty
\frac{du\,W'(u)}{1 + \lambda^2\nu^2 \lt(\ln u + \mathcal{N}\rt)^2} \rt\}
\end{equation}
See Figures~\ref{comparing} and~\ref{steepening}.

For a hyperbolic tangent step
\begin{equation}\label{hstep}
f(x) = \tanh x
\end{equation}
we have
\begin{equation}
\tilde{f}(p) = \frac{1}{2i\sinh(\pi p/2)}
\end{equation}
Therefore Eq.~(\ref{Fourier}) gives
\begin{equation}
\mathcal{P}_{\mathcal{R}_\mathrm{c}}
= \frac{V_0 e^{\lambda^2\mathcal{N}}}{12\pi^2\lambda^2}
\lt\{ 1 + \left(\alpha - \frac{5}{6} \right) \lambda^2
- 2 \lt(\frac{A\nu}{\lambda}\rt)
\int_0^\infty \frac{p}{\sinh(\pi p/2)}\,\mathrm{Re}\lt[
Q(ip\lambda\nu)\,e^{ip\lambda\nu\mathcal{N}}\rt] dp \rt\}
\end{equation}
Alternatively Eq.~(\ref{iPex}) gives
\begin{equation}
\mathcal{P}_{\mathcal{R}_\mathrm{c}}
= \frac{V_0 e^{\lambda^2\mathcal{N}}}{12\pi^2\lambda^2} \lt\{ 1
+ \lt( \alpha - \frac{5}{6} \rt) \lambda^2
+ 2 \lt(\frac{A\nu}{\lambda}\rt) \int_0^\infty
\frac{du\,W'(u)}{\cosh^2\lt[\lambda\nu \lt(\ln u + \mathcal{N}\rt)\rt]} \rt\}
\end{equation}

For a Gaussian integral step
\begin{equation}\label{Gstep}
f(x) = \int_0^x e^{-t^2} dt
\end{equation}
we have
\begin{equation}
\tilde{f}(p) = \frac{1}{2\sqrt{\pi}\,ip} e^{-p^2/4}
\end{equation}
Therefore Eq.~(\ref{Fourier}) gives
\begin{equation}
\mathcal{P}_{\mathcal{R}_\mathrm{c}}
= \frac{V_0 e^{\lambda^2\mathcal{N}}}{12\pi^2\lambda^2}
\lt\{ 1 + \left(\alpha - \frac{5}{6} \right) \lambda^2
- \frac{2}{\sqrt{\pi}\,} \lt(\frac{A\nu}{\lambda}\rt)
\int_0^\infty e^{-p^2/4}\,\mathrm{Re}\lt[Q(ip\lambda\nu)\,
e^{ip\lambda\nu\mathcal{N}}\rt] dp \rt\}
\end{equation}
Alternatively Eq.~(\ref{iPex}) gives
\begin{equation}
\mathcal{P}_{\mathcal{R}_\mathrm{c}}
= \frac{V_0 e^{\lambda^2\mathcal{N}}}{12\pi^2\lambda^2} \lt\{ 1
+ \lt( \alpha - \frac{5}{6} \rt) \lambda^2
+ 2 \lt(\frac{A\nu}{\lambda}\rt) \int_0^\infty du\,W'(u)\,
\exp\lt[-\lambda^2\nu^2 \lt(\ln u + \mathcal{N}\rt)^2\rt] \rt\}
\end{equation}

\subsection{Sharp step}

A sharp step is given by taking the large $|\lambda\nu|$ limit
of a smooth step such as those above.
Letting $s = \lambda\nu(\ln u + \mathcal{N})$
we can write the integral in Eq.~(\ref{iPex}) as
\begin{equation}
\int_0^\infty du\,W'(u)\,f'(\lambda\nu\mathcal{N} + \lambda\nu\ln u)
= \frac{1}{|\lambda\nu|} \int_{-\infty}^\infty ds\
e^{-\mathcal{N}}e^\frac{s}{\lambda\nu}\,
W'\lt(e^{-\mathcal{N}}e^\frac{s}{\lambda\nu}\rt) f'(s)
\end{equation}
For large $|\lambda\nu|$ one can show
\begin{equation}
e^{-\mathcal{N}}e^\frac{s}{\lambda\nu}\,
W'\lt(e^{-\mathcal{N}}e^\frac{s}{\lambda\nu}\rt)
\sim e^{-\mathcal{N}}\,W'\lt(e^{-\mathcal{N}}\rt)
\cos\lt(\frac{2s}{\lambda\nu}e^{-\mathcal{N}}\rt)
+ 3 \sin\lt(2e^{-\mathcal{N}}\rt) \sin\lt(\frac{2s}{\lambda\nu}e^{-\mathcal{N}}\rt)
\end{equation}
with errors of order $s/(\lambda\nu)$.
Therefore, if $f'(s)$ goes to zero sufficiently rapidly as $|s| \rightarrow \infty$,
Eq.~(\ref{iPex}) gives
\begin{eqnarray}\label{sharp}
\lefteqn{
\mathcal{P}_{\mathcal{R}_\mathrm{c}}
= \frac{V_0 e^{\lambda^2\mathcal{N}}}{12\pi^2\lambda^2} \lt\{ 1
+ \lt( \alpha - \frac{5}{6} \rt) \lambda^2
+ 2 \lt(\frac{A\nu}{\lambda}\rt) \frac{1}{|\lambda\nu|}
\rt. } \\ \nn && \lt. \times
\lt[ e^{-\mathcal{N}}\,W'\lt(e^{-\mathcal{N}}\rt) \int_{-\infty}^\infty ds\,f'(s)\,
\cos\lt(\frac{2s}{\lambda\nu}e^{-\mathcal{N}}\rt)
+ 3 \sin\lt(2e^{-\mathcal{N}}\rt) \int_{-\infty}^\infty ds\,f'(s)\,
\sin\lt(\frac{2s}{\lambda\nu}e^{-\mathcal{N}}\rt) \rt] \rt\}
\end{eqnarray}
Taking the limit $|\lambda\nu| \rightarrow \infty$ with $\mathcal{N}$ fixed gives
\begin{equation}\label{sharplimit}
\mathcal{P}_{\mathcal{R}_\mathrm{c}}
= \frac{V_0 e^{\lambda^2\mathcal{N}}}{12\pi^2\lambda^2} \lt\{ 1
+ \lt( \alpha - \frac{5}{6} \rt) \lambda^2
+ 2 \lt(\frac{A\nu}{\lambda}\rt)
\lt[\frac{f(\infty)-f(-\infty)}{|\lambda\nu|}\rt]
e^{-\mathcal{N}}\,W'\lt(e^{-\mathcal{N}}\rt) \rt\}
\end{equation}
where
\begin{eqnarray}
e^{-\mathcal{N}}\,W'(e^{-\mathcal{N}})
& = &
- 3 \lt( 1 - 3 e^{2\mathcal{N}} \rt) \cos\lt(2e^{-\mathcal{N}}\rt)
+ \frac{3}{2} \lt( 5 - 3 e^{2\mathcal{N}} \rt)
e^\mathcal{N} \sin\lt(2e^{-\mathcal{N}}\rt) \\
& \sim  & \frac{4}{5} e^{-2\mathcal{N}}
\ \ \mbox{as}\ \ \mathcal{N} \rightarrow \infty \\
& \sim  & - 3 \cos\lt(2e^{-\mathcal{N}}\rt)
\ \ \mbox{as}\ \ \mathcal{N} \rightarrow - \infty
\end{eqnarray}
This could have been obtained more simply by substituting $f(x) = \theta(x)$
into Eq.~(\ref{iPex}), but that method would miss the damping of the
oscillations that occurs for $\mathcal{N} \lesssim - \ln |\lambda\nu|$.
See Figure~\ref{steepening}.
For example, for an arctanget step, Eq.~(\ref{astep}), Eq.~(\ref{sharp}) gives
\begin{equation}
\mathcal{P}_{\mathcal{R}_\mathrm{c}}
= \frac{V_0 e^{\lambda^2\mathcal{N}}}{12\pi^2\lambda^2} \lt\{ 1
+ \lt( \alpha - \frac{5}{6} \rt) \lambda^2
+ 2 \lt(\frac{A\nu}{\lambda}\rt) \frac{\pi}{|\lambda\nu|}
\exp\lt(-\frac{2e^{-\mathcal{N}}}{|\lambda\nu|}\rt)
e^{-\mathcal{N}}\,W'\lt(e^{-\mathcal{N}}\rt) \rt\}
\end{equation}
for a hyperbolic tangent step, Eq.~(\ref{hstep}), Eq.~(\ref{sharp}) gives
\begin{equation}
\mathcal{P}_{\mathcal{R}_\mathrm{c}}
= \frac{V_0 e^{\lambda^2\mathcal{N}}}{12\pi^2\lambda^2} \lt\{ 1
+ \lt( \alpha - \frac{5}{6} \rt) \lambda^2
+ 2 \lt(\frac{A\nu}{\lambda}\rt) \frac{2}{|\lambda\nu|}
\frac{\pi e^{-\mathcal{N}}/(\lambda\nu)}{\sinh\lt[\pi e^{-\mathcal{N}}/(\lambda\nu)\rt]}
e^{-\mathcal{N}}\,W'\lt(e^{-\mathcal{N}}\rt) \rt\}
\end{equation}
and for a Gaussian integral step, Eq.~(\ref{Gstep}), Eq.~(\ref{sharp}) gives
\begin{equation}
\mathcal{P}_{\mathcal{R}_\mathrm{c}}
= \frac{V_0 e^{\lambda^2\mathcal{N}}}{12\pi^2\lambda^2} \lt\{ 1
+ \lt( \alpha - \frac{5}{6} \rt) \lambda^2
+ 2 \lt(\frac{A\nu}{\lambda}\rt) \frac{\sqrt{\pi}}{|\lambda\nu|}
\exp\lt[-\lt(\frac{e^{-\mathcal{N}}}{\lambda\nu}\rt)^2\rt]
e^{-\mathcal{N}}\,W'\lt(e^{-\mathcal{N}}\rt) \rt\}
\end{equation}

\subsection{Bump or dip}

For an arctangent derivative bump or dip
\begin{equation}
f(x) = \frac{1}{1+x^2}
\end{equation}
we have
\begin{equation}
\tilde{f}(p) = \frac{1}{2} e^{-|p|}
\end{equation}
Therefore Eq.~(\ref{Fourier}) gives
\begin{equation}
\mathcal{P}_{\mathcal{R}_\mathrm{c}}
= \frac{V_0 e^{\lambda^2\mathcal{N}}}{12\pi^2\lambda^2}
\lt\{ 1 + \left(\alpha - \frac{5}{6} \right) \lambda^2
- 2 \lt(\frac{A\nu}{\lambda}\rt)
\int_0^\infty e^{-p}\,\mathrm{Re}\lt[ip
\,Q(ip\lambda\nu)\,e^{ip\lambda\nu\mathcal{N}}\rt] dp \rt\}
\end{equation}
Alternatively Eq.~(\ref{iPex}) gives
\begin{equation}
\mathcal{P}_{\mathcal{R}_\mathrm{c}}
= \frac{V_0 e^{\lambda^2\mathcal{N}}}{12\pi^2\lambda^2} \lt\{ 1
+ \lt( \alpha - \frac{5}{6} \rt) \lambda^2
- 4 \lt(\frac{A\nu}{\lambda}\rt) \lambda\nu \int_0^\infty
\frac{du\,W'(u) \lt(\ln u + \mathcal{N}\rt)}
{\lt[ 1 + \lambda^2\nu^2 \lt(\ln u + \mathcal{N}\rt)^2 \rt]^2} \rt\}
\end{equation}

For a hyperbolic tangent derivative bump or dip
\begin{equation}
f(x) = \frac{1}{\cosh^2 x}
\end{equation}
we have
\begin{equation}
\tilde{f}(p) = \frac{p}{2\sinh(\pi p/2)}
\end{equation}
Therefore Eq.~(\ref{Fourier}) gives
\begin{equation}
\mathcal{P}_{\mathcal{R}_\mathrm{c}}
= \frac{V_0 e^{\lambda^2\mathcal{N}}}{12\pi^2\lambda^2}
\lt\{ 1 + \left(\alpha - \frac{5}{6} \right) \lambda^2
- 2 \lt(\frac{A\nu}{\lambda}\rt)
\int_0^\infty \frac{p}{\sinh(\pi p/2)}\,\mathrm{Re}\lt[ip
\,Q(ip\lambda\nu)\,e^{ip\lambda\nu\mathcal{N}}\rt] dp \rt\}
\end{equation}
Alternatively Eq.~(\ref{iPex}) gives
\begin{equation}
\mathcal{P}_{\mathcal{R}_\mathrm{c}}
= \frac{V_0 e^{\lambda^2\mathcal{N}}}{12\pi^2\lambda^2} \lt\{ 1
+ \lt( \alpha - \frac{5}{6} \rt) \lambda^2
- 4 \lt(\frac{A\nu}{\lambda}\rt) \int_0^\infty
\frac{du\,W'(u)\,\sinh\lt[\lambda\nu \lt(\ln u + \mathcal{N}\rt)\rt]}
{\cosh^3\lt[\lambda\nu \lt(\ln u + \mathcal{N}\rt)\rt]} \rt\}
\end{equation}

For a Gaussian bump or dip
\begin{equation}
f(x) = e^{-x^2}
\end{equation}
we have
\begin{equation}
\tilde{f}(p) = \frac{1}{2\sqrt{\pi}\,} e^{-p^2/4}
\end{equation}
Therefore Eq.~(\ref{Fourier}) gives
\begin{equation}
\mathcal{P}_{\mathcal{R}_\mathrm{c}}
= \frac{V_0 e^{\lambda^2\mathcal{N}}}{12\pi^2\lambda^2}
\lt\{ 1 + \left(\alpha - \frac{5}{6} \right) \lambda^2
- \frac{2}{\sqrt{\pi}\,} \lt(\frac{A\nu}{\lambda}\rt)
\int_0^\infty e^{-p^2/4}\,\mathrm{Re}\lt[ip\,Q(ip\lambda\nu)\,
e^{ip\lambda\nu\mathcal{N}}\rt] dp \rt\}
\end{equation}
Alternatively Eq.~(\ref{iPex}) gives
\begin{equation}
\mathcal{P}_{\mathcal{R}_\mathrm{c}}
= \frac{V_0 e^{\lambda^2\mathcal{N}}}{12\pi^2\lambda^2} \lt\{ 1
+ \lt( \alpha - \frac{5}{6} \rt) \lambda^2
- 4 \lt(\frac{A\nu}{\lambda}\rt) \lambda\nu \int_0^\infty du\,W'(u)
\lt(\ln u + \mathcal{N}\rt)
\exp\lt[-\lambda^2\nu^2 \lt(\ln u + \mathcal{N}\rt)^2\rt] \rt\}
\end{equation}

\setcounter{section}{0}
\renewcommand{\thesection}{\Alph{section}}
\section{Appendices}

\subsection{Notation}

In this appendix I summarize the notation used in this paper.
\begin{equation}
d\eta \equiv \frac{dt}{a}
\,,\ \
x \equiv -k\eta
\end{equation}
\begin{equation}
\varphi \equiv a\left(\delta\phi-\frac{\dot\phi}{H}\mathcal{R}\right)
\,,\ \
y \equiv \sqrt{2k}\, \varphi_k
\,,\ \
y_0(x) \equiv \left(1 + \frac{i}{x}\right)e^{ix}
\end{equation}
\begin{equation}
z \equiv \frac{a\dot\phi}{H}
\,,\ \
f(\ln x) \equiv xz
\,,\ \
g(\ln x) \equiv \frac{f''(\ln x)-3f'(\ln x)}{f(\ln x)}
\end{equation}
\begin{equation}
\epsilon \equiv \frac{1}{2}\left(\frac{\dot{\phi}}{H}\right)^2
\,,\ \
\delta_n \equiv \frac{1}{H^n\dot{\phi}}\left(\frac{d}{dt}\right)^n\dot\phi
\,,\ \
\mathcal{U} \equiv \left(\frac{V'}{V}\right)^2
\,,\ \
\mathcal{V}_n \equiv \left(\frac{V'}{V}\right)^{n-1} \frac{V^{(n+1)}}{V}
\end{equation}
\begin{equation}
\alpha \equiv 2 - \ln 2 - \gamma \simeq 0.729637
\end{equation}
where $\gamma$ is the Euler-Mascheroni constant.
\begin{equation}
\sum_{n=0}^\infty d_n(x_\star)\,\nu^n \equiv D_\star(\nu)
\equiv (2x_\star)^\nu \cos\left(\frac{\pi\nu}{2}\right) \frac{\Gamma(2-\nu)}{1+\nu}
\end{equation}
\begin{equation}
\sum_{n=0}^\infty q_n(x_\star)\,\nu^n \equiv Q_\star(\nu)
\equiv (2x_\star)^{-\nu}\cos\left(\frac{\pi\nu}{2}\right)\frac{3\Gamma(2+\nu)}{(1-\nu)(3-\nu)}
\end{equation}
\begin{equation}
D(\nu) \equiv \lt.D_\star(\nu)\rt|_{x_\star=1}
\,,\ \
Q(\nu) \equiv \lt.Q_\star(\nu)\rt|_{x_\star=1}
\end{equation}
\begin{equation}
\omega(x) \equiv \frac{\sin(2x)}{x} - \cos(2x)
\end{equation}
\begin{equation}
W(x) \equiv
\frac{3\sin(2x)}{2x^3} - \frac{3\cos(2x)}{x^2} - \frac{3\sin(2x)}{2x}
\end{equation}

\subsection{Slow-roll parameters}

In this paper I assume the general slow-roll approximation
\begin{eqnarray}\label{my1}
\epsilon & = & \mathcal{O}(\xi)
\\ \label{my2}
\delta_n & = & \mathcal{O}(\xi)
\end{eqnarray}
for some small perturbation parameter $\xi$,
where
\begin{eqnarray}
\epsilon \equiv \frac{1}{2} \lt(\frac{\dot\phi}{H}\rt)^2
\\
\delta_n \equiv \frac{1}{H^n\dot\phi} \lt(\frac{d}{dt}\rt)^n \dot\phi
\end{eqnarray}
The slow-roll parameters are related by
\begin{eqnarray}\label{edot}
\frac{d\epsilon}{d\ln a} & = & 2 \lt(\epsilon + \delta_1\rt) \epsilon
\\ \label{ddot}
\frac{d\delta_n}{d\ln a}
& = & \delta_{n+1} + \lt(n\epsilon - \delta_1\rt) \delta_n
\end{eqnarray}
Note that
\begin{equation}\label{econst}
\frac{d\ln\epsilon}{d\ln a} = 2\lt(\epsilon+\delta_1\rt) = \mathcal{O}(\xi)
\end{equation}
and so $\epsilon$ is approximately constant in the slow-roll
approximation.
The standard slow-roll approximation assumes the $\delta_n$'s are also
approximately constant, but this is not true in general and I will not
assume it.

\subsection{Slow-roll parameters in terms of the inflaton potential}

To express $H$, $\epsilon$ and the $\delta_n$'s in terms of
$V$, $\mathcal{U}$ and the $\mathcal{V}_n$'s, we use
\begin{equation}
3H^2 = \frac{1}{2}\dot{\phi}^2 + V(\phi)
\end{equation}
its derivatives and $\dot{H}=-\dot{\phi}^2/2$ to give
\begin{eqnarray}
\frac{V}{3H^2} & = & 1 - \frac{1}{3}\epsilon \label{VH} \\
\frac{V'}{3H^2} & = & - \frac{\dot{\phi}}{H} \left( 1 + \frac{1}{3}\delta_1 \right)
\label{VpH} \\
\frac{V''}{3H^2} & = & \epsilon - \delta_1 - \frac{1}{3}\delta_2
\end{eqnarray}
Therefore
\begin{eqnarray}\label{srV1}
\frac{H^4}{\dot{\phi}^2} & = & \frac{V^3}{3{V'}^2}
\lt[ 1 + \epsilon + \frac{2}{3} \delta_1 + \mathcal{O}\lt(\xi^2\rt) \rt]
\\ \label{Vpe}
\mathcal{U} \ \equiv \ \left(\frac{V'}{V}\right)^2
& = & 2\epsilon + \mathcal{O}\lt(\xi^2\rt)
\\ \label{Vppe}
\mathcal{V}_1 \ \equiv \ \frac{V''}{V}
& = & \epsilon - \delta_1 - \frac{1}{3}\delta_2 + \mathcal{O}\lt(\xi^2\rt)
\end{eqnarray}
Now
\begin{equation}\label{dHdt}
\frac{d}{Hdt} = - \left(\frac{1-\frac{1}{3}\epsilon}{1+\frac{1}{3}\delta_1}\right)
\frac{V'}{V} \frac{d}{d\phi}
\end{equation}
Therefore, using Eqs.~(\ref{edot}) and~(\ref{ddot}),
\begin{equation}\label{Vp}
\mathcal{V}_n \equiv \left(\frac{V'}{V}\right)^{n-1} \frac{V^{(n+1)}}{V}
= (-1)^n \left( \delta_n + \frac{1}{3}\delta_{n+1} \right)
+ \mathcal{O}\lt(\xi^2\rt) \ \ \ \mbox{for $n\geq2$}
\end{equation}
Therefore
\begin{eqnarray}\label{d1V}
\delta_1 & = & \frac{1}{2}\mathcal{U}
- \sum_{j=0}^\infty \left(\frac{1}{3}\right)^j \mathcal{V}_{j+1}
+ \mathcal{O}\lt(\xi^2\rt) \\
\delta_n & = & (-1)^n \sum_{j=0}^\infty \left(\frac{1}{3}\right)^j \mathcal{V}_{n+j}
+ \mathcal{O}\lt(\xi^2\rt) \ \ \ \mbox{for $n\geq2$}
\end{eqnarray}

\subsection{$x$, $f$ and $g$ in terms of the slow-roll parameters
and the inflaton potential}

\begin{equation}\label{x}
x = -k\eta = -k \int \frac{dt}{a}
= \frac{k}{aH} \lt[ 1 + \epsilon + \mathcal{O}\lt(\xi^2\rt) \rt]
\end{equation}
Therefore
\begin{equation}\label{fsrV}
\frac{1}{f^2} = \frac{1}{x^2 z^2} = \frac{H^4}{k^2\dot{\phi}^2}
\lt[ 1 - 2 \epsilon + \mathcal{O}\lt(\xi^2\rt) \rt]
= \frac{V^3}{3k^2{V'}^2}
\lt[ 1 - \frac{1}{2} \mathcal{U} + \frac{2}{3} \delta_1
+ \mathcal{O}\lt(\xi^2\rt) \rt]
\end{equation}
\begin{equation}\label{fp}
\frac{f'}{f} = \frac{d\ln f}{d\ln x} = - 2 \epsilon - \delta_1 + \mathcal{O}\lt(\xi^2\rt)
= - \mathcal{U} - \delta_1 + \mathcal{O}\lt(\xi^2\rt)
\end{equation}
and for $n\geq2$
\begin{equation}
\frac{f^{(n)}}{f} = \lt(\frac{f^{(n-1)}}{f}\rt)' + \mathcal{O}\lt(\xi^2\rt)
= \lt(\frac{f'}{f}\rt)^{(n-1)} + \mathcal{O}\lt(\xi^2\rt)
= (-1)^n \delta_n + \mathcal{O}\lt(\xi^2\rt)
\end{equation}
Therefore
\begin{equation}\label{gsrV}
g = \frac{f''}{f} - 3 \frac{f'}{f}
= 6 \epsilon + 3 \delta_1 + \delta_2 + \mathcal{O}\lt(\xi^2\rt)
= \frac{9}{2} \mathcal{U} - 3 \mathcal{V}_1 + \mathcal{O}\lt(\xi^2\rt)
\end{equation}

Now from Eq.~(\ref{fexp})
\begin{equation}
\frac{f_n}{f_0} = \frac{f^{(n)}(\ln x_\star)}{f(\ln x_\star)}
\end{equation}
therefore
\begin{eqnarray}
\frac{f_1}{f_0} & = & - 2\epsilon_\star - {\delta_1}_\star + \mathcal{O}\lt(\xi^2\rt)
\\
\frac{f_n}{f_0} & = & (-1)^n {\delta_n}_\star + \mathcal{O}\lt(\xi^2\rt)
\ \ \ \mbox{for $n\geq 2$}
\end{eqnarray}
Eq.~(\ref{g}) gives
\begin{equation}
g_n = - 3 \frac{f_n}{f_0} + \frac{f_{n+1}}{f_0} + \mathcal{O}\lt(\xi^2\rt)
\end{equation}
therefore
\begin{eqnarray}\label{g1}
g_1 & = & 6 \epsilon_\star + 3 {\delta_1}_\star + {\delta_2}_\star
+ \mathcal{O}\lt(\xi^2\rt)
\ = \ \frac{9}{2} \mathcal{U}_\star - 3 {\mathcal{V}_1}_\star
+ \mathcal{O}\lt(\xi^2\rt)
\\ \label{gn}
g_n & = & (-1)^{n+1} \left( 3 {\delta_n}_\star + {\delta_{n+1}}_\star \right)
+ \mathcal{O}\lt(\xi^2\rt)
\ = \ - 3 {\mathcal{V}_n}_\star + \mathcal{O}\lt(\xi^2\rt)
\ \ \ \mbox{for $n\geq 2$}
\end{eqnarray}

\subsection{Other useful formulae}

\begin{equation}\label{aHk}
\frac{d X|_{aH=k}}{d\ln k}
= \left. \frac{\partial X}{\partial\ln (aH)} \right|_{aH=k}
+ \left. \frac{\partial X}{\partial\ln k} \right|_{aH=k}
= \left. \frac{\dot{X}}{(1-\epsilon)H} \right|_{aH=k}
+ \left. \frac{\partial X}{\partial\ln k} \right|_{aH=k}
\end{equation}

\subsection{Relationship between the series formulae and the integral formulae}

The series and integral formulae can be related using the following formulae
\begin{eqnarray}
\frac{D_\star(\nu)-1}{\nu} & = & - \int_0^\infty \frac{du}{u}
\lt(\frac{x_\star}{u}\rt)^\nu \lt[ \omega(u) - \theta(x_\star-u) \rt]
\\
\omega(x) - \theta(x_\star-x)
& = & - \frac{1}{2\pi i} \int_{-i\infty}^{i\infty} d\nu
\lt(\frac{x}{x_\star}\rt)^\nu \lt[\frac{D_\star(\nu)-1}{\nu}\rt]
\\
\frac{Q_\star(\nu)-1}{\nu} & = & \int_0^\infty \frac{du}{u}
\lt(\frac{u}{x_\star}\rt)^\nu \lt[ W(u) - \theta(x_\star-u) \rt]
\\ \label{QWp}
Q_\star(\nu) & = & - \int_0^\infty du \lt(\frac{u}{x_\star}\rt)^\nu W'(u)
\\
W(x) - \theta(x_\star-x)
& = & \frac{1}{2\pi i} \int_{-i\infty}^{i\infty} d\nu
\lt(\frac{x_\star}{x}\rt)^\nu \lt[\frac{Q_\star(\nu)-1}{\nu}\rt]
\\
x\,W'(x) & = & - \frac{1}{2\pi i} \int_{-i\infty}^{i\infty} d\nu
\lt(\frac{x_\star}{x}\rt)^\nu Q_\star(\nu)
\end{eqnarray}
For example, Eqs.~(\ref{sPVU}) and~(\ref{iPV2}) are related by Eq.~(\ref{QWp})
as follows
\begin{eqnarray}
\lt[ Q_\star\lt(\frac{V'}{V}\frac{d}{d\phi}\rt) \ln\mathcal{U} \rt]_\star
& = & \lt[ Q_\star\lt(\frac{d}{d\ln x}\rt) \ln\mathcal{U} \rt]_\star
+ \mathcal{O}\lt(\xi^2\rt) \\
& = & \lt\{ - \int_0^\infty du\,W'(u) \exp\lt[\ln\lt(\frac{u}{x_\star}\rt)
\frac{d}{d\ln x}\rt] \ln\mathcal{U} \rt\}_\star
+ \mathcal{O}\lt(\xi^2\rt) \\
& = & - \int_0^\infty du\,W'(u) \lt.\ln\mathcal{U}\rt|_{aH=\frac{k}{u}}
+ \mathcal{O}\lt(\xi^2\rt)
\end{eqnarray}

\subsection{Mathematical formulae}
\label{Amath}

In this appendix I give some mathematical formulae used in this paper.
\begin{equation}
\psi(\nu) \equiv \frac{d}{d\nu} \ln \Gamma(\nu)
\end{equation}
\begin{equation}
\psi(1) = - \gamma \simeq - 0.577215665
\end{equation}
\begin{equation}
\psi^{(n)}(1) = (-1)^{n+1} n! \, \zeta(n+1) \ \ \ \mbox{for $n\geq1$}
\end{equation}
\begin{equation}
\zeta(2)=\pi^2/6
\,,\ \
\zeta(3)\simeq1.202056903
\,,\ \
\zeta(4)=\pi^4/90
\end{equation}
\begin{equation}
\zeta(n) = \frac{2^{n-1}B_n\pi^n}{n!} \ \ \ \mbox{for $n$ even, $n\geq2$}
\end{equation}
where $B_2=1/6$, $B_4=1/30$, \ldots \ are the Bernoulli numbers.

\subsection{Calculating the $d_n$'s}
\label{Adn}

Appendix~\ref{Amath} contains some mathematical formulae used in this section.
The $d_n$'s are given by the generating function
\begin{equation}
\sum_{n=0}^\infty d_n(x_\star)\,\nu^n \equiv D_\star(\nu)
\equiv (2x_\star)^\nu \cos\left(\frac{\pi\nu}{2}\right) \frac{\Gamma(2-\nu)}{1+\nu}
\equiv \mathrm{Re} \left[ e^{E_\star(\nu)} \right]
\end{equation}
where
\begin{equation}
E_\star(\nu) = \ln\Gamma(1-\nu) + \ln(1-\nu) - \ln(1+\nu)
+ \left[\ln(2x_\star)+\frac{i\pi}{2}\right]\nu
\end{equation}
Therefore
\begin{equation}
d_n(x_\star) = \frac{1}{n!} D^{(n)}_\star(0) = \mathrm{Re} \left[
\sum_{j_1,\ldots,j_n\geq0}^{j_1+\ldots+nj_n=n} \frac{1}{j_1! \dots j_n!}
\left(\frac{E^{(1)}_\star(0)}{1!}\right)^{j_1} \ldots
\left(\frac{E^{(n)}_\star(0)}{n!}\right)^{j_n} \right]
\end{equation}
where
\begin{equation}
E^{(n)}_\star(0) = \left\{ \begin{array}{ll}
- \alpha_\star + \frac{i\pi}{2} & \mbox{for $n=1$} \\
(n-1)! \, \zeta(n) & \mbox{for $n$ even, $n\geq2$} \\
(n-1)! \, \left[\zeta(n)-2\right] & \mbox{for $n$ odd, $n\geq3$}
\end{array} \right.
\end{equation}
and
\begin{equation}
\alpha_\star
= 2 - \ln 2 - \gamma - \ln x_\star
= \alpha - \ln x_\star
\end{equation}
Some explicit values are
\begin{eqnarray}
d_0 & = & 1 \\
d_1 & = & -\alpha_\star \\
d_2 & = & \frac{\alpha_\star^2}{2} - \frac{\pi^2}{24} \\
d_3 & = & -\frac{\alpha_\star^3}{6} + \frac{\alpha_\star\pi^2}{24}
- \frac{1}{3}\left[2-\zeta(3)\right] \\
d_4 & = & \frac{\alpha_\star^4}{24} - \frac{\alpha_\star^2\pi^2}{48}
+ \frac{\alpha_\star}{3}\left[2-\zeta(3)\right] - \frac{\pi^4}{640}
\end{eqnarray}
Numerical values for $x_\star=1$ are $d_1 \simeq -0.729637$,
$d_2 \simeq -0.145048$, $d_3 \simeq -0.030669$, $d_4 \simeq -0.055787$.
$d_1$ agrees with Ref.~\cite{foc} and $d_2$ with Ref.~\cite{Jinook}.

\subsection{Calculating the $q_n$'s}
\label{Aqn}

The $q_n$'s are given by the generating function
\begin{equation}
\sum_{n=0}^\infty q_n(x_\star) \nu^n \equiv Q_\star(\nu)
\equiv (2x_\star)^{-\nu} \cos\left(\frac{\pi\nu}{2}\right)
\frac{3\Gamma(2+\nu)}{(1-\nu)(3-\nu)}
\end{equation}
They are related to the $d_n$'s by
\begin{equation}
d_n = (-1)^n \left( q_n - \frac{1}{3} q_{n-1} \right)
\end{equation}
Therefore
\begin{equation}
q_n = (-1)^n d_n + \frac{1}{3} q_{n-1}
= \frac{1}{3^n} \sum_{j=0}^n d_j (-3)^j
\end{equation}
Some explicit values are
\begin{eqnarray}
q_0 & = & 1 \\
q_1 & = & \beta_\star \\
q_2 & = & \frac{\beta_\star^2}{2} - \frac{\pi^2}{24} + \frac{1}{18} \\
q_3 & = & \frac{\beta_\star^3}{6} - \frac{\pi^2\beta_\star}{24} + \frac{\beta_\star}{18}
+ \frac{1}{3}\left[2-\zeta(3)\right] + \frac{1}{81} \\
q_4 & = & \frac{\beta_\star^4}{24} - \frac{\pi^2\beta_\star^2}{48} + \frac{\beta_\star^2}{36}
+ \frac{\beta_\star}{3}\left[2-\zeta(3)\right] + \frac{\beta_\star}{81}
- \frac{\pi^4}{640} - \frac{\pi^2}{432 } + \frac{1}{216}
\end{eqnarray}
where $ \beta_\star \equiv \alpha_\star + \frac{1}{3}
= \alpha + \frac{1}{3} - \ln x_\star$.
Numerical values for $x_\star = 1$ are $q_1 \simeq 1.062970$, $q_2 \simeq 0.209275$,
$q_3 \simeq 0.100428$, $q_4 \simeq -0.022311$.
$q_1$ agrees with Ref.~\cite{foc} and $q_2$ with Ref.~\cite{Jinook}.

\subsection*{Acknowledgements}
I thank Scott Dodelson, Jin-Ook Gong, Salman Habib, Gerard Jungman, Hyun-Chul Lee and Carmen Molina-Paris for valuable discussions, and the SF01 Cosmology Summer Workshop and the Fermilab Theoretical Astrophysics Group for hospitality.
This work was supported in part by Brain Korea 21 and KRF grant 2000-015-DP0080.

\end{document}